\documentclass[conference,compsoc]{IEEEtran}

\usepackage[nocompress]{cite}
\usepackage{xurl} 
\usepackage{graphicx}
\usepackage[caption=false,font=footnotesize]{subfig}
\usepackage{booktabs}
\usepackage{multirow}
\usepackage{xcolor}
\usepackage{enumitem}
\usepackage{hyperref}

\usepackage[most]{tcolorbox}
\usepackage{fontawesome5}
\hypersetup{hidelinks}

\newif\ifdraftsummaries
\draftsummariesfalse

\title{Copilot or Culprit? How AI Is Changing Open-Source Bug Reporting and Maintenance}

\author{
\IEEEauthorblockN{
Yinuo Liu\IEEEauthorrefmark{1},
Yunlong Lyu\IEEEauthorrefmark{1},
Zichao Luo\IEEEauthorrefmark{2},
Tian Dong\IEEEauthorrefmark{1},
Zian Luo\IEEEauthorrefmark{3},
Peng Chen\IEEEauthorrefmark{4},
Hao Chen\IEEEauthorrefmark{1},
}
\IEEEauthorblockA{
\IEEEauthorrefmark{1}The University of Hong Kong, Hong Kong, Hong Kong\\
\IEEEauthorrefmark{2}Huazhong University of Science and Technology, China\\
\IEEEauthorrefmark{3}Xi'an Jiaotong University, China\\
\IEEEauthorrefmark{4}Independent Researcher\\
}
}

\begin{document}
\bstctlcite{BSTcontrol}
\clubpenalty=10000
\widowpenalty=10000
\brokenpenalty=10000
\maketitle
\begin{abstract}
AI-generated bug reports have drawn criticism for unreliable claims and faulty reproducers that burden maintainers.  However, as LLM capabilities improve, some maintainers have begun to express more positive views on AI use in bug reporting and maintenance. Understanding this evolution requires examining whether AI-involved reports are becoming more reliable, which uses of AI are associated with valid findings, and how maintainers investigate these reports and act on them. 
To this end, we conduct the first large-scale study of 120,918 bug reports from 101 open-source projects between January 2024 and September 2026, tracing AI involvement from discovery and reporting through validation and resolution. Our findings show that AI-involved reports become more complete over time, with a growing share of reports confirmed as valid.
In bug reporting, AI for reproduction and diagnosis are associated with higher validity, whereas report writing is not.
Extends into maintenance, maintainers increasingly use AI to investigate code and recommend fixes. However, they devote more effort to AI-assisted reports of human-discovered bugs than to reports of AI-discovered bugs.
These findings highlight both AI’s growing participation in bug investigation and the continuing need for verification and follow-up. 
Drawing on these findings, we provide practical guidance for turning AI-generated findings into verifiable reports: emphasizing reproducibility, verification, and accountability. We followed this guidance when submitting 53 reports on AI-discovered bugs, of which 69.81\% were accepted by maintainers as of writing.
\end{abstract}

\section{Introduction}
\label{sec:introduction}

\begin{figure}[t]
    \centering
    \includegraphics[width=\columnwidth]{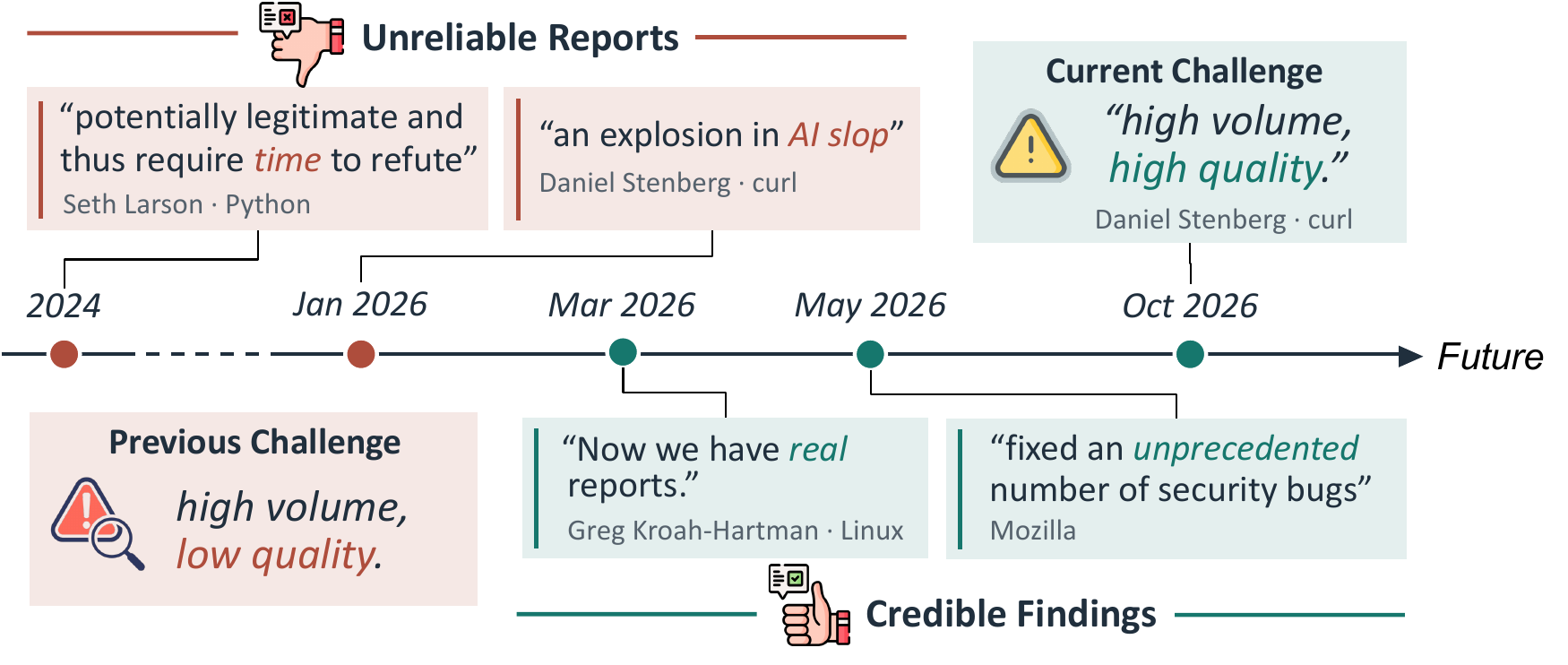}
    \caption{Evolving challenges of AI-generated reports.}
    \vspace{-4mm}
    \label{fig:timeline}
\end{figure}

AI-generated bug reports have raised concerns about unreliable claims, false positives, and increasing maintenance demands. As early as 2024, maintainers warned that seemingly credible but incorrect reports required substantial effort to investigate and refute~\cite{larson2024slop}. Similarly, Log4j maintainers reported that the influx of AI-generated security reports consumed substantial volunteer effort and disrupted regular development~\cite{karwasz2026log4j}. As illustrated in \autoref{fig:timeline}, the growing volume of unreliable reports placed considerable pressure on maintainers across open-source projects. In response to mounting submission and review burdens, some organizations suspended vulnerability reporting programs. In January 2026, curl ended its bug bounty program after a surge of low-quality submissions, with the proportion of reports confirmed as vulnerabilities falling below 5\%~\cite{stenberg2026curl}. Organizations such as HackerOne and Google also suspended certain vulnerability submission channels amid growing pressures from AI-accelerated reports~\cite{vijayan2026ibb,google2026osspause,gatlan2026google}.

However, recent advances in Large Language Models (LLMs) have produced increasingly useful findings, leading some maintainers to reconsider their views on AI-generated reports. Mozilla reported validating and fixing AI-discovered vulnerabilities supported by reproducible test cases~\cite{grinstead2026firefox}. Linux maintainer Greg Kroah-Hartman observed a shift from low-quality AI submissions to \textit{``real reports''}~\cite{vaughannichols2026linux}. Similarly, curl reported that the proportion of confirmed vulnerabilities had recovered to around 16\% by April 2026~\cite{stenberg2026chaos}. By October 2026, curl's maintainer described the challenge as \textit{``high volume, high quality''}, reflecting growing demands for validating and resolving genuine findings~\cite{stenberg2026ignore}. As summarized in \autoref{fig:timeline}, these developments suggest that the challenges of AI-involved bug reporting are shifting from filtering unreliable claims to validating and resolving genuine defects, motivating a systematic study of AI involvement throughout reporting and maintenance.

Prior work has largely focused on the negative impacts of AI-generated contributions, documenting low-quality submissions and their burden on maintainers~\cite{baltes2026slop,afroz2026slop}, examining project policies~\cite{yang2026governance,chen2026aipolicy}, and measuring pull request volume and merge rates~\cite{afroz2026slop}. Other studies evaluate tools for improving reports, detecting invalid submissions, and supporting automated repair~\cite{Bo2024chatbr,akyol2026improbr,zheng2025reviewers,Khatib2026what}. However, as AI-generated findings become increasingly useful, existing research offers a limited understanding of how AI is changing bug reporting and maintenance. This motivates us to systematically study how AI participates in bug discovery and reporting (RQ1), how report reliability evolves over time and which factors are associated with validity (RQ2), and how maintainers investigate, handle, and use AI when responding to these reports (RQ3).

\noindent\textbf{Our work.}~ To this end, we conduct the first large-scale systematic study of AI involvement at the individual bug report level, analyzing 120,918 reports from 101 open-source projects between January 2024 and September 2026. Three key challenges emerge when conducting such a study. First, tracing AI involvement requires determining when, by whom, and for which tasks AI was used, based on fragmented information in issue discussions. Second, quantifying report reliability and maintainer burden requires translating ambiguous discussions and behaviors into well-defined, measurable indicators. Third, substantial differences across projects, time periods, and AI roles complicate meaningful comparisons. To address these challenges, we develop a measurement framework covering AI roles and tasks, report completeness and validity, and maintainer activities and AI use. We establish the measurement criteria through manual analysis and reviewer discussions, then apply LLM-based annotation with human validation. To account for project and temporal differences, we compare reports within the same project and quarter and examine differences across AI roles and factors associated with validity.

\noindent\textbf{Main Findings.}~
Our study reveals five key findings:

\begin{itemize}[topsep=2pt, itemsep=2pt, parsep=0pt]
    \item AI involvement in bug reporting is growing rapidly, from 0.7\% in 2024 to 8.3\% in 2026-Q3. Beyond assisting human reporters, AI increasingly discovers bugs through source code review (Section~\ref{sec:ai-use}).

    \item AI-involved reports tend to be more complete, but are not necessarily more valid. Nevertheless, both information completeness and validity among decided reports improve over time, by 5.41 and 4.89 percentage points per quarter, respectively (Section~\ref{sec:reliability}).

    \item In bug reporting, AI-assisted reproduction and diagnosis are associated with higher validity (OR = 2.25 and 1.42), whereas report writing is not. However, AI-generated materials still require verification: maintainers reject 23.53\% of AI-generated reproducers, compared with 5.14\% of others (Sections~\ref{subsec:validity-factors} and~\ref{subsec:maintainer-material}).

    \item AI-assisted reports of human-discovered bugs receive more maintainer actions (+0.46 per report), driving deeper investigation, whereas reports of AI-discovered bugs receive fewer (-0.93), receiving less attention (Sections~\ref{subsec:interactions} and~\ref{subsec:actions}). 

    \item AI use also extends into maintenance, where maintainers increasingly use it for code investigation and fix recommendations. AI-involved maintainer replies rose from 0.07\% in 2024 to 3.40\% in 2026 (Section~\ref{subsec:maintainer-ai}).
\end{itemize}

\noindent\textbf{Recommendations.}~
Drawing on these findings, we provide practical recommendations for reporters and maintainers. For reporters, we emphasize reproducibility, verification of AI-generated claims, and human accountability for submission and follow-up. For maintainers, we recommend adapting reporting policies to growing AI involvement, standardizing report formats, screening common errors before deeper investigation, and preserving AI-assisted investigation results for subsequent review. To demonstrate how these recommendations can be applied, we followed the reporting principles when submitting 53 reports of AI-discovered bugs, of which 69.81\% received valid maintainer verdicts, reaching 94.87\% among reports with definitive verdicts. These results illustrate a practical approach to turning AI-generated findings into verifiable bug reports.

\noindent\textbf{Contributions.}~This study makes three main contributions:
\begin{itemize}[topsep=2pt, itemsep=2pt, parsep=0pt]
    \item \textbf{Lifecycle Measurement.}
    We develop a framework tracing AI involvement from bug discovery and reporting to maintainer validation and resolution, covering 120,918 reports from 101 projects.

    \item \textbf{Empirical Insights.}
    We reveal how report reliability evolves, which AI uses are associated with valid findings, and how maintainer practices vary across AI roles.

    \item \textbf{Practical Guidance.}
    We provide actionable recommendations for bug reporting and maintenance, demonstrating their practical value by applying our reporting principles to 53 AI-discovered bugs, achieving a 69.81\% maintainer-confirmed validity rate versus 25.51\% in our dataset.
\end{itemize}

\section{Background and Related Works}
\label{sec:background}

\noindent\textbf{AI in Bug Reporting.}~
Generative AI can participate in bug reporting by discovering potential defects or assisting humans with problems they have already encountered. For bug discovery, LLMs can identify potential defects by analyzing source code or generating tests that expose failures. Code analysis approaches use LLMs to inspect suspicious code, perform static analysis, and audit repositories for vulnerabilities~\cite{grinstead2026firefox,anthropic2026mythos,Li2025iris,Guo2025repoaudit}. Alternatively, LLMs support fuzzing by generating test inputs~\cite{Xia2024fuzz4all,Deng2023titanfuzz,Deng2024fuzzgpt,Meng2024chatafl}, constructing fuzz drivers~\cite{Lyu2024promptfuzz}, and automating fuzzing workflows through agents~\cite{Lyu2026fuzzagent}. Beyond discovery, LLMs can help reporters diagnose failures, construct reproducers, and document their findings. Existing tools improve report completeness and clarity~\cite{Bo2024chatbr,akyol2026improbr}, generate structured reports~\cite{Acharya2025reports,Choi2025agentreport}, and incorporate execution context into descriptions and reproduction steps~\cite{Saha2026bugscribe}. Together, these approaches extend AI's role from identifying potential defects to preparing technical materials submitted to maintainers. However, generated interpretations may remain unverified and should be distinguished from observed failures.

\noindent\textbf{AI in Report Handling.}~
After submission, maintainers typically handle bug reports through triage, validation, and resolution~\cite{Torun2026past}. Generative AI can support each stage of this process. During triage, maintainers organize incoming reports and determine how they should be handled. LLMs can classify issues~\cite{Colavito2025issueclassification}, identify duplicate reports~\cite{Laney2025duplicate}, and recommend suitable developers~\cite{Kiashemshaki2025triaging}. Beyond triage, maintainers must determine whether reported failures represent genuine defects. LLM-based tools support this process by identifying potentially invalid reports~\cite{zheng2025reviewers}, generating tests to reproduce failures~\cite{Kang2023libro}, and replaying failures from bug descriptions~\cite{Feng2024adbgpt,Wang2024rebl}. For confirmed bugs, coding agents can investigate source code, locate faults, generate patches, and test proposed fixes~\cite{Yang2024sweagent,Zhang2024autocoderover,Bouzenia2025repairagent,Xia2025agentless}. AI also supports code review by checking coding practices~\cite{Vijayvergiya2024autocommenter}. Together, these approaches extend AI's role across bug investigation and resolution. As reports may already contain AI-generated technical materials, maintainers must also verify their correctness before relying on them.

\noindent\textbf{Empirical Studies of AI Use in OSS.}~
Prior empirical studies examine AI involvement in OSS from three perspectives. First, research investigates AI adoption and its impact on OSS communities. Some studies measure AI use across projects and development activities~\cite{Tufano2026selfadmitted,Robbes2026agentic}, while others document low-quality contributions, maintainer burden, and project governance~\cite{baltes2026slop,afroz2026slop,yang2026governance,chen2026aipolicy}. These studies reveal changes in OSS communities but offer limited insights into how AI usage relates to the reliability of bug reports. Second, as AI agents increasingly contribute code, researchers examine how maintainers review their contributions. Using datasets such as AIDev~\cite{Li2026aidev}, studies investigate rejection reasons~\cite{Abujadallah2026rejection,Ehsani2026failures}, reviewer interventions~\cite{Peralta2026outcomes,Gao2026autopilot}, and review practices~\cite{Duma2026reviews,Njoku2026agents}. While these studies characterize human interactions with AI contributions, they focus on pull requests rather than bug reports. Third, research extends beyond code contributions to examine AI assistance in issue resolution. Existing studies investigate how developers use AI during debugging and problem solving~\cite{Hao2024conversations,Das2025issuetracker}, and identify factors associated with successful resolution~\cite{Ehsani2026issueresolution}. However, they focus on developer interactions rather than how maintainers validate and handle AI-involved reports. Together, these studies provide insights into AI adoption, code contributions, and issue resolution, but offer limited longitudinal evidence connecting AI usage in bug reporting with report reliability and maintainer practices. Our study addresses this gap by tracing individual bug reports throughout reporting and maintenance.

\section{Study Design}
\label{sec:methodology}

We first introduce our research questions and measurement framework, covering AI usage, report reliability, and maintainer practices. We then describe the data collection and coding procedures, followed by the statistical methods used to analyze the results.

\begin{figure*}[t]
    \centering
    \includegraphics[width=\textwidth]{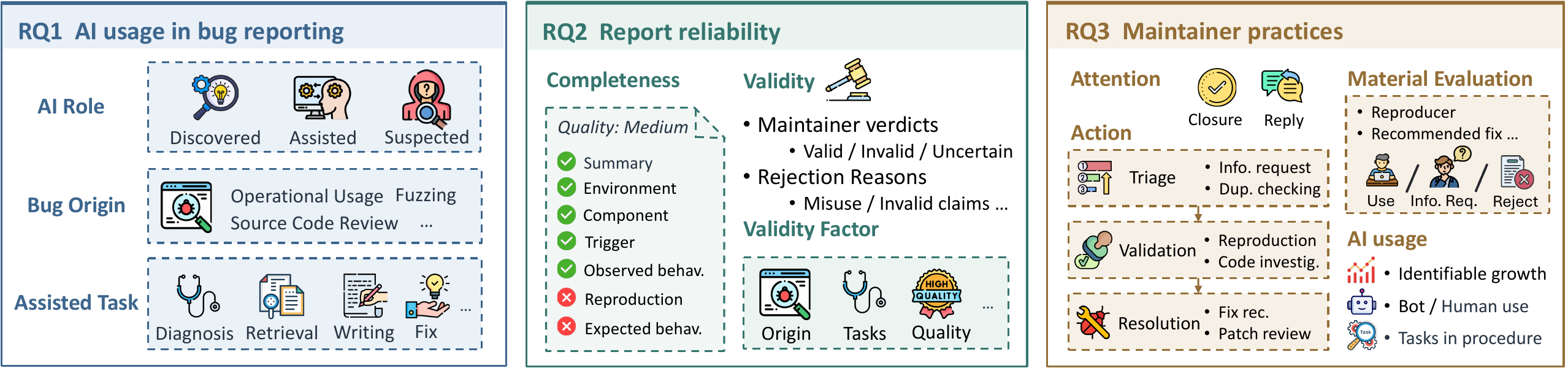}
    \caption{Measurement framework for AI usage, report reliability, and maintainer practices.}
    \label{fig:measurement_framework}
\end{figure*}
\vspace{-2mm}
\subsection{Research Questions}
\label{subsec:research_questions}

We investigate how generative AI is changing bug reporting in open-source projects through three research questions covering AI's roles in reporting, the reliability of AI-involved reports, and its influence on maintainer practices.

\noindent\textbf{RQ1: How is generative AI used in bug reporting?}
This question focuses on the adoption and roles of AI in bug reporting. It covers how widely AI is used, how its involvement evolves over time, and where AI-involved reports originate. It also considers how different AI-supported tasks are combined throughout the reporting process.

\noindent\textbf{RQ2: How reliable are AI-involved reports?}
This question examines whether AI-involved reports provide sufficient information for investigation, whether maintainers confirm the reported bugs, and how both evolve over time. It also investigates why maintainers reject reports and which factors are associated with validity.

\noindent\textbf{RQ3: How does AI influence the handling of bug reports?}
This question explores how AI-involved reports alter maintainer workflows. Specifically, it considers how maintainers allocate initial attention, how they distribute effort across triage, validation, and resolution stages, how they evaluate explicitly AI-generated materials, and how they adopt AI tools in their own practices.

\subsection{Measurement Framework}
\label{subsec:measurement}

As illustrated in \autoref{fig:measurement_framework}, our measurement framework covers three aspects: AI usage in bug reporting, report reliability, and maintainer practices. Using public bug reports, we trace AI involvement from discovery and reporting to validation and resolution.

\noindent\textbf{AI Usage in Bug Reporting (RQ1).}
We examine AI involvement through three dimensions: AI role, bug discovery method, and AI-assisted tasks.
For \emph{AI role}, we distinguish four categories based on how AI participates in reporting: (1) \emph{AI-discovered}, where AI first identifies a potential bug; (2) \emph{AI-assisted}, where AI assists with a bug initially encountered by humans; (3) \emph{suspected}, where maintainers suspect AI generation without confirmation; and (4) \emph{Other}, where AI involvement is absent or unknown. The first three constitute \textit{AI-involved} reports.
For \emph{bug discovery method}, we classify how the problem was initially identified into source code review, static analysis, fuzzing, dynamic analysis, testing, build or installation, and operational usage, with unknown used when the method cannot be determined. This separates the discovery process from AI's role in it.
For \emph{AI-assisted tasks}, we identify six non-exclusive activities: problem diagnosis, debugging support, reproduction and testing, fix recommendation, information retrieval, and report writing. We also examine task co-occurrence to capture how AI supports multiple activities within a report.

\noindent\textbf{Report Reliability (RQ2).}
We assess report reliability through two dimensions: \emph{information completeness}, which measures whether reports provide sufficient information for investigation, and \emph{technical validity}, which reflects whether maintainers confirm the reported bugs. Drawing on prior studies of bug report quality~\cite{Zimmermann2010,Soltani2020}, we examine seven elements in initial reports: summary, affected component, triggering conditions, reproduction steps, observed behavior, expected behavior, and environment. Based on missing or insufficient elements, we classify reports into \emph{high}, \emph{medium}, and \emph{low} completeness levels. Because complete reports may still describe invalid bugs, we assess validity separately using maintainer responses, classifying reports as \emph{valid}, \emph{invalid}, or \emph{uncertain}. To understand why reports are rejected, we examine six non-exclusive rejection reasons based on prior research on invalid bug reports~\cite{Laiq2023} and our manual coding. These include misuse, expected behavior, unreproducibility, incorrect claims, out-of-scope issues, and others. Finally, we examine how discovery methods, AI-assisted tasks, and report completeness, measured through overall quality levels and individual missing elements, are associated with validity.

\noindent\textbf{Maintainer Practices (RQ3).}
To understand how maintainers handle AI-involved reports, we examine their engagement, activities, evaluation of report materials, and use of AI. Prior studies have examined developer interactions and the stages of issue resolution~\cite{Huang2019Questions,Saha2025Resolution}. Building on this work and our manual coding, we define four measurement dimensions. For \emph{maintainer attention}, we measure closure rates and reply counts to capture initial engagement. For \emph{maintainer actions}, we classify actions into three stages: (1) \emph{triage}, covering information requests and duplicate checking; (2) \emph{validation}, covering reproduction and testing and code and API investigation; and (3) \emph{resolution}, covering fix recommendations and patch review and testing. For \emph{material evaluation}, we examine whether maintainers use, request supplementation for, or reject reproducers, technical claims, and proposed fixes, distinguishing the evaluation of individual materials from overall bug validity. Finally, for \emph{maintainer-side AI use}, we distinguish AI bots from human maintainers using AI, tracking adoption and the activities supported. We compare these measures across AI roles to characterize differences in maintainer practices.

\subsection{Data Collection}
\label{subsec:data-collection}

\noindent\textbf{Project Selection.}
We study 101 open-source projects across eight widely used programming languages: C, C++, Python, Java, JavaScript, TypeScript, Go, and Rust to cover diverse software ecosystems. For each language, we identify the 100 most-starred GitHub repositories to ensure that our candidate pool includes popular and established projects~\cite{Borges2018Stars}, excluding non-software repositories such as curated lists and tutorials. We retain projects with at least one new issue in 2026 and between 500 and 4,000 issues from January 2024 to September 2026, ensuring sufficient activity while reducing the risk that high-volume projects disproportionately influence the results~\cite{Gorostidi2024Sampling}. Finally, we obtain 376 eligible repositories and randomly select 100 from them. We additionally include the Linux kernel, one of the most well-known open-source projects, resulting in 101 projects. The projects span diverse domains, including systems software, networking, data processing, machine learning, AI-agent frameworks, and development tools.

\noindent\textbf{Issue Collection.}
We include issues opened from January 1, 2024 through September 30, 2026 (UTC), together with their initial reports, follow-up discussions, and metadata such as labels, status, and timestamps. We retain closure and comment records available as of October 24, 2026. Corpus-profile counts and monthly trends use the calendar window ending September 30, while maintainer-confirmed verdicts and analyses of all replies use the full snapshot. For closure and reply comparisons, we require at least 90 days of observation.
Since not all issues describe software defects~\cite{Fan2020chaff}, we use an LLM to identify bug reports based on issue content. We include reports describing potential defects, unexpected behavior, or security vulnerabilities, regardless of subsequent maintainer verdicts. Feature requests, support questions, and documentation suggestions are excluded. To validate the classification, three reviewers manually examined a sample of 500 LLM-labeled reports, and all agreed with the assigned labels.

\noindent\textbf{Dataset Overview.}
Our initial corpus contains 171,819 records across 101 projects, with 656,555 follow-up comments from 148,774 participant accounts. After screening, we identify 120,918 bug reports, accompanied by a total of 503,596 follow-up comments collected through October 24, 2026. These reports involved 118,830 participant accounts during the study period. \autoref{tab:sample-profile} summarizes the dataset.

\begin{table}[t]
\centering
\caption{Overview of our dataset.}
\label{tab:sample-profile}
\small
\setlength{\tabcolsep}{4pt}
\resizebox{0.9\columnwidth}{!}{%
\begin{tabular}{@{}lrrc@{}}
\toprule
Statistics & Mean & Median & Distribution \\
\midrule
\# of Stars & 60,923 & 50,500 & \raisebox{-0.45\height}{\includegraphics[width=1in]{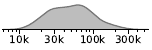}} \\
\# of Issues & 1,701 & 1,523 & \raisebox{-0.45\height}{\includegraphics[width=1in]{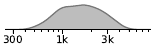}} \\
\# of Bug Reports & 1,197 & 964 & \raisebox{-0.45\height}{\includegraphics[width=1in]{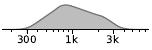}} \\
\# of Comments & 6,501 & 5,331 & \raisebox{-0.45\height}{\includegraphics[width=1in]{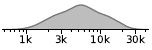}} \\
\midrule
\multicolumn{4}{@{}p{\columnwidth}@{}}{\footnotesize C++ (27), Python (23), C (19), Java (10), JS (6), TS (6), Go (6), Rust (4)} \\
\bottomrule
\end{tabular}}
\end{table}

\subsection{Coding Procedure}

\noindent\textbf{Hybrid Coding.}
We adopt a hybrid coding approach combining inductive and deductive coding~\cite{Fereday2006hybrid} to develop our measurement framework. We use the Linux kernel, OpenSSL, and OpenCV as pilot projects and manually examine randomly sampled bug
reports and their discussions. During the inductive phase, we identify recurring patterns in bug discovery methods, reporter AI tasks, report elements, quality levels, rejection reasons, and maintainer actions. We organize these findings into an initial codebook with clear category definitions, coding rules, and representative examples. During the deductive phase, we apply the codebook to systematically classify the remaining reports, using LLMs to support large-scale annotation. Finally, three reviewers with software engineering backgrounds review the codebook and classified examples, discuss ambiguous cases such as identifying bug discovery methods, and refine the definitions until consensus is reached.

\noindent\textbf{LLM Annotation.}
Given the scale of our dataset, we use GLM~5.3~\cite{Zai2026glm53} to annotate bug reports based on the developed codebook. We divide the annotation into three tasks: bug report identification, report characterization, and maintainer response analysis. For bug report identification, the LLM distinguishes bug reports from feature requests, support questions, and other non-bug issues. For report characterization, it examines the initial report and relevant reporter discussions to identify bug origin, AI involvement, AI tasks, and information quality. For maintainer response analysis, it examines maintainer replies to identify report validity, rejection reasons, maintainer actions, and AI usage. We provide the LLM with category definitions, coding rules, and a structured output schema. Appendix~\ref{app:definitions} defines the study measures in the order of the research questions and clarifies their coding boundaries. For labels requiring explicit evidence, the LLM is instructed to extract supporting text from the issue conversations.

\noindent\textbf{Validation.}
To assess the reliability of our annotations, we independently re-annotated the full dataset using DeepSeek-V4.1-Flash-1017~\cite{DeepSeek2026V41Flash1017}, following the same codebook and prompt template as GLM~5.3. We then compared the outputs of both models across annotation tasks to identify disagreements. Three reviewers with software engineering backgrounds examined these cases against the original bug reports and issue discussions, considering the contextual information and coding rules to determine the appropriate labels. All disagreements were resolved through manual inspection and discussion to maintain consistency across the final annotations.

\subsection{Statistical Analysis}
\label{subsec:statistical-analysis}
We summarize AI adoption, report characteristics, and maintainer activities with descriptive statistics. Our main inferential analyses use two methods.

\noindent\textbf{Group comparisons.}
We compare AI-involved and other reports within the same project and quarter. We average differences in proportions or means across strata containing both groups, weighted by the number of AI-involved reports in each stratum. We resample projects 5,000 times to estimate cluster-bootstrap standard errors and percentile 95\% confidence intervals~\cite{Cameron2015cluster}. We calculate two-sided normal-approximation $p$-values using the estimated difference divided by its bootstrap standard error. Comparisons for individual AI roles follow the same procedure.

\noindent\textbf{Factors associated with validity.}
We use Firth logistic regression~\cite{Firth1993bias} on reports with definitive verdicts, controlling for AI role and project and quarter fixed effects. We estimate project-clustered standard errors~\cite{Cameron2015cluster} and use Wald tests for individual coefficients and joint Wald tests for groups of factors. We report odds ratios and 95\% confidence intervals; model specifications are detailed in Section~\ref{subsec:validity-factors}.

We also assess exploratory linear quarterly trends within AI-involved reports using project and role fixed effects and project-clustered standard errors. Trend specifications and sensitivity analyses are described in Appendix~\ref{app:temporal-analysis}.

Some analyses use different tests to accommodate the data. For the small sample of explicitly AI-generated materials, we use quarter-stratified exact conditional tests~\cite{Agresti1992exact}, as detailed in Section~\ref{subsec:maintainer-material}.

We apply Benjamini--Hochberg correction~\cite{Benjamini1995fdr} to the group comparisons, regression Wald tests, and exact conditional tests, treating adjusted $p$-values ($q$) below 0.05 as significant. Correction is performed within each test family: overall and role-specific group comparisons are corrected separately, as are individual regression coefficients and factor-group tests. Quarterly trend tests are corrected across six outcomes within each model specification and sample. Exact tests are grouped by analysis, as specified in Section~\ref{subsec:maintainer-material}. We report unadjusted 95\% confidence intervals.

\section{How Is AI Used in Bug Reporting?}
\label{sec:ai-use}
To understand how AI is integrated into bug reporting, we examine its adoption and workflow activities. First, we track the scale and role distribution of AI involvement over time. Second, we analyze bug discovery sources to understand how AI-driven and human-driven workflows differ. Finally, focusing on AI-assisted reports, we examine AI tasks and their combinations to characterize how reporters use AI after encountering a failure.

\subsection{Scale of AI Use}
\label{subsec:scale}

\begin{figure}[t]
\centering
\includegraphics[width=\columnwidth]{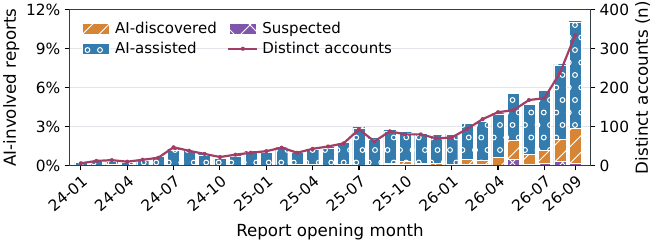}
\caption{Monthly AI-involved reports by role and distinct reporting accounts from January 2024 to September 2026.}
\label{fig:ai-growth}
\end{figure}

We identify 3,000 AI-involved reports among 120,918 bug reports (2.5\%), submitted by 2,316 accounts across 101 projects. Of these, 2,508 are AI-assisted, 396 AI-discovered, and 96 suspected. Python repositories contribute 47.5\% of AI-involved reports, compared with 23.8\% of all reports.

\noindent\textbf{AI involvement in bug reporting has grown rapidly.}
As shown in \autoref{fig:ai-growth}, AI-involved reports increased from 292 in 2024 to 769 in 2025 and 1,939 in the first nine months of 2026. Their share of all bug reports rose from 0.7\% to 1.9\% and 8.3\% in 2026-Q3, reaching 11.1\% in September. Meanwhile, overall reporting volume remained relatively stable, indicating growing AI adoption.

\noindent\textbf{AI is increasingly used for bug discovery, beyond assisting human reporters.}
AI-assisted reports account for 83.6\% of AI-involved reports, but AI-discovered reports grew from 23 in 2025 to 372 in the first nine months of 2026. Their share rose from 3.0\% to 19.2\%, reaching 23.8\% in September. This suggests that AI increasingly identifies potential bugs rather than only assisting with human-discovered problems.

\noindent\textbf{AI-involved reports come from a growing number of accounts and projects.}
The number of reporting accounts increased from 259 in 2024 to 708 in 2025 and 1,385 in the first nine months of 2026. Projects receiving AI-involved reports grew from 55 to 87 and 99. Meanwhile, the share contributed by the five most active projects fell from 71.2\% to 22.3\%. These trends indicate broader participation and a less concentrated distribution across projects.

\begin{table}[t]
\centering
\caption{Distribution of bug discovery methods across AI roles. AI-disc.: AI-discovered; AI-assist.: AI-assisted; Suspect.: suspected AI involvement.}

\label{tab:ai-origin}
\footnotesize
\setlength{\tabcolsep}{3pt}
\begin{tabular}{@{}lrrrr@{}}
\toprule
\textbf{Discovery Method} & \textbf{AI-disc.} & \textbf{AI-assist.} & \textbf{Suspect.} & \textbf{Other} \\
\midrule
Source Code Review   & \textbf{51.52\%} & \underline{6.66\%} & \underline{11.46\%} & 5.62\% \\
Static Analysis      & 5.05\%  & 0.44\%  & 0.00\%  & 0.31\% \\
Fuzzing              & 4.29\%  & 0.44\%  & 0.00\%  & 0.67\% \\
Dynamic Analysis     & 0.25\%  & 0.24\%  & 2.08\%  & 0.60\% \\
Testing              & 3.79\%  & 4.23\%  & 5.21\%  & 4.70\% \\
Build / Installation & 0.51\%  & 3.39\%  & 4.17\%  & \underline{5.65\%} \\
Operational Usage    & \underline{13.89\%} & \textbf{70.06\%} & \textbf{60.42\%} & \textbf{65.71\%} \\
Unknown              & 20.71\% & 14.55\% & 16.67\% & 16.74\% \\
\bottomrule
\end{tabular}
\end{table}

\subsection{Discovery Sources}
\label{subsec:discovery-sources}

While Section~\ref{subsec:scale} shows a growing split between AI-assisted and AI-discovered reports, role labels alone do not reveal where AI enters the reporting process. We therefore classify how bugs were initially discovered into seven categories: Source Code Review, Static Analysis, Fuzzing, Dynamic Analysis, Testing, Build / Installation, and Operational Usage. Unspecified cases are labeled Unknown. \autoref{tab:ai-origin} compares their distributions across AI roles.

\noindent\textbf{AI-discovered reports primarily originate from source code review.}
Source Code Review accounts for 51.5\% of AI-discovered reports, compared with 6.7\% of AI-assisted reports and 5.6\% of other reports. Static Analysis and Fuzzing account for another 5.1\% and 4.3\% of AI-discovered reports, respectively. In contrast, only 13.9\% originate from Operational Usage. These results link AI-driven discovery mainly to code inspection, with fewer reports arising from normal software use.

\noindent\textbf{AI-assisted reports largely follow human discovery patterns.}
Operational Usage accounts for 70.1\% of AI-assisted reports, close to 65.7\% of other reports. Their Source Code Review shares are also similar, at 6.7\% and 5.6\%, respectively. This suggests that AI commonly assists with problems already encountered by reporters.

\begin{table*}[!t]
\centering
\caption{Information completeness by AI role.}
\label{tab:info-completeness}
\renewcommand{\arraystretch}{1.12}
\setlength{\tabcolsep}{7pt}
\resizebox{\textwidth}{!}{%
\begin{tabular}{@{}l ccc ccccccc c@{}}
\toprule
\multirow{2}{*}{\textbf{AI Role}}
& \multicolumn{3}{c}{\textbf{Quality}}
& \multicolumn{7}{c}{\textbf{Missing Elements}}
& \multirow{2}{*}{\shortstack{\textbf{Avg.}\\\textbf{Missing}}} \\
\cmidrule(lr){2-4}
\cmidrule(lr){5-11}
& \textbf{High} & \textbf{Med.} & \textbf{Low}
& \textbf{Summary} & \textbf{Comp.} & \textbf{Trigger}
& \textbf{Repro.} & \textbf{Observed} & \textbf{Expected}
& \textbf{Env.} & \\
\midrule
\textbf{Other}
& 33.77\% & 51.91\% & 14.32\%
& 9.69\% & 7.60\% & 22.09\%
& 37.72\% & 16.39\% & 42.77\%
& 29.19\% & 1.65 \\
\midrule
\textbf{AI-disc.}
& 56.31\% & 37.12\% & 6.57\%
& 1.52\% & 1.52\% & 7.32\%
& 18.69\% & 9.60\% & 16.67\%
& 30.05\% & 0.85 \\
\textbf{AI-assist.}
& 50.00\% & 42.86\% & 7.14\%
& 3.11\% & 2.23\% & 9.81\%
& 26.36\% & 6.22\% & 27.19\%
& 21.33\% & 0.96 \\
\textbf{Suspect.}
& 45.83\% & 42.71\% & 11.46\%
& 4.17\% & 3.12\% & 13.54\%
& 30.21\% & 9.38\% & 29.17\%
& 28.12\% & 1.18 \\
\midrule
\textbf{AI-invol.}
& 50.70\% & 42.10\% & 7.20\%
& 2.93\% & 2.17\% & 9.60\%
& 25.47\% & 6.77\% & 25.87\%
& 22.70\% & 0.95 \\
\textbf{Adj. $\Delta$}
& $\mathbf{+8.72}^{***}$
& $-4.04^{*}$
& $\mathbf{-4.68}^{***}$
& $-4.44^{***}$
& $-4.19^{***}$
& $\mathbf{-8.16}^{***}$
& $\mathbf{-8.42}^{***}$
& $-7.72^{***}$
& $\mathbf{-8.93}^{***}$
& $-5.21^{**}$
& $\mathbf{-0.47}^{***}$ \\
\bottomrule
\end{tabular}%
}

\parbox{\textwidth}{\footnotesize
\vspace{1mm}
\textit{Note.} Significance after BH correction: $^{*}q<0.05$, $^{**}q<0.01$, $^{***}q<0.001$. \textbf{AI-invol.}: all AI-involved reports;  \textbf{Med.}: medium; \textbf{Comp.}: component; \textbf{Repro.}: reproduction steps; \textbf{Env.}: environment. Adjusted differences ($\Delta$) compare AI-involved and other reports within the same project and quarter.
}
\end{table*}

\subsection{AI Tasks in Bug Reporting}
\label{subsec:ai-tasks}
To understand how reporters use AI after encountering failures, we examine the 2,084 AI-assisted reports with identifiable tasks. We classify these activities into six non-exclusive categories: Problem Diagnosis, Debugging Support, Reproduction and Testing, Fix Recommendation, Information Retrieval, and Report Writing.

\noindent\textbf{AI primarily supports problem diagnosis and fix recommendations.}~
Problem Diagnosis is the most common task, appearing in 55.7\% of reports, followed by Fix Recommendation at 42.0\% and Report Writing at 32.9\%. Reproduction and Testing, Information Retrieval, and Debugging Support account for 12.7\%, 10.3\%, and 8.8\%, respectively. These findings show that AI helps reporters understand failures and propose fixes, rather than simply prepare report text.

\begin{figure}[t]
\centering
\includegraphics[width=0.75\columnwidth]{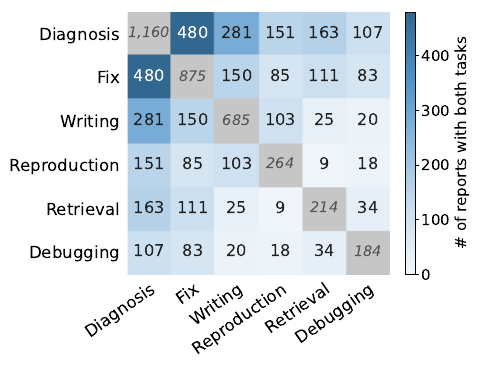}

\caption{Co-occurrence of AI tasks in assisted bug reports. \textbf{Diagnosis}: problem diagnosis; \textbf{Fix}: fix recommendation; \textbf{Writing}: report writing; \textbf{Reproduction}: reproduction and testing; \textbf{Retrieval}: information retrieval; \textbf{Debugging}: debugging support.}

\label{fig:task-cooccur}
\end{figure}

\noindent\textbf{AI often participates in multiple reporting tasks.}
Among the 2,084 reports, 42.0\% involve at least two AI tasks. \autoref{fig:task-cooccur} shows that Problem Diagnosis is the most frequent partner for every other task. The most common combination is Problem Diagnosis and Fix Recommendation, appearing in 480 reports, followed by Problem Diagnosis and Report Writing in 281 reports. These patterns suggest that reporters use AI across several activities, including understanding failures, proposing fixes, and preparing reports.

\begin{tcolorbox}[
    enhanced,
    colback=black!6,
    colframe=black,
    boxrule=0.5pt,
    arc=3mm,
    left=5pt,
    right=5pt,
    top=4pt,
    bottom=4pt,
    before skip=8pt,
    after skip=8pt
]
\noindent
\faLightbulb[regular]\quad
\textbf{Key Takeaways (RQ1):}~
AI use is rapidly expanding across reporters, mainly assisting with failures encountered during software use. In these reports, AI supports diagnosis and fix recommendations, often combining multiple tasks. Meanwhile, AI-discovered reports are growing rapidly primarily through source code review, extending AI's role from investigating failures to discovering potential defects.
\end{tcolorbox}

\section{How Reliable Are AI-Involved Reports?}
\label{sec:reliability}
Although AI increasingly supports bug discovery and reporting (Section~\ref{sec:ai-use}), whether the resulting reports are reliable remains unclear. To answer RQ2, we first assess information completeness, maintainer-confirmed validity, and their temporal evolution. We then examine why maintainers reject certain reports, and finally model how discovery methods, AI tasks, and report attributes relate to validity.

\subsection{Information Completeness}
\label{subsec:information-completeness}

Starting with whether reports provide sufficient context, we compare AI-involved and other reports across three quality levels and seven information elements. We account for project and temporal differences by comparing reports within 550 project-quarter groups. We calculate weighted differences using the number of AI-involved reports as weights. We bootstrap projects 5,000 times to estimate 95\% confidence intervals and apply Benjamini--Hochberg correction to eleven comparisons.

\noindent\textbf{AI-involved reports have higher information quality.}~
As shown in \autoref{tab:info-completeness}, 50.70\% of AI-involved reports are high quality, versus 33.77\% of other reports. The gap remains significant after adjusting for project and temporal differences (8.72 percentage points, $q<0.001$). AI-involved reports are also 4.68 percentage points less likely to be low quality, while AI-discovered reports have the highest high-quality rate (56.31\%). This may be because LLMs can transform brief user inputs into structured descriptions, while automated discovery workflows incorporate execution logs and test inputs into report generation.

\noindent\textbf{AI-involved reports contain fewer missing elements.}~
AI-involved reports have significantly lower missing rates across all seven elements, averaging 0.47 fewer missing elements per report after adjustment. The largest differences appear in expected behavior, reproduction steps, and triggering conditions, with reductions of 8.93, 8.42, and 8.16 percentage points, respectively. While human reporters may leave expectations implicit or omit intermediate steps, LLMs can infer expected behavior from error descriptions and articulate failure conditions step by step. However, approximately one quarter of AI-involved reports still lack reproduction steps or expected behavior. This gap remains when users rely on AI only to translate or polish brief notes without providing underlying runtime details.

\begin{figure}[t]
\centering
\includegraphics[width=0.9\columnwidth]{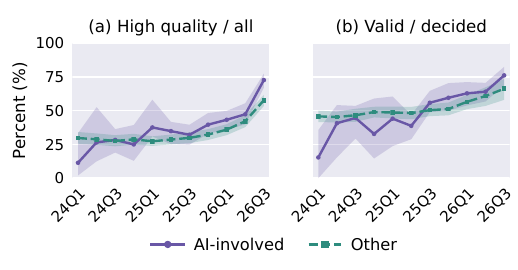}
\caption{Quarterly report quality: (a) high-quality share among all reports; (b) valid share among decided reports. Shading shows 95\% project-cluster bootstrap confidence intervals.}
\label{fig:report-quality-timeline}
\end{figure}

\noindent\textbf{Information completeness improves over time.}~
\autoref{fig:report-quality-timeline}(a) shows an increasing share of high-quality reports among all AI-involved reports across opening cohorts. After controlling for project and AI role, this share increases by 5.41 percentage points per quarter ($q<0.001$). This trend persists after excluding 2026Q3. This steady increase reflects both improved capability in newer models and a broader shift from standalone web chats to IDE-integrated and agentic tools that automatically capture local context and issue templates.

\subsection{Technical Validity}
\label{subsec:validity}

Despite the higher completeness found in Section~\ref{subsec:information-completeness}, information completeness does not necessarily imply technical validity. We classify maintainer verdicts as valid, invalid, or uncertain, and measure validity among decided reports. We follow the project-quarter comparison procedure in Section~\ref{subsec:information-completeness}, correcting across four verdict measures.

\noindent\textbf{AI-involved reports show no significant difference in maintainer-confirmed validity.}~
After adjustment, AI-involved and other reports do not differ significantly in invalid rate or validity among decided reports, with both differences at $-0.37$ percentage points ($q\geq0.44$). AI-discovered reports have a lower valid rate by 10.92 points ($q=0.033$), while 62.37\% remain uncertain. Their invalid rate is not significantly higher. Detailed verdict rates and role comparisons appear in Appendix \autoref{tab:validity}. This suggests that LLMs improve surface presentation but not necessarily bug diagnosis. AI-discovered reports often involve subtle or theoretical flaws requiring further investigation, leaving many unconfirmed rather than rejected.

\noindent\textbf{Validity among decided reports improves over time.}~ \autoref{fig:report-quality-timeline}(b) shows an increasing valid share among AI-involved reports with definitive verdicts. After controlling for project and AI role, this share increases by 4.89 percentage points per quarter ($q<0.001$). The trend persists after excluding 2026Q3. However, the uncertain share rises from 43.15\% in 2024 to 52.19\% in the first nine months of 2026, so this improvement describes reports with definitive verdicts. This divergence likely reflects two factors: stronger reasoning and tool-use capabilities help newer models filter obvious false positives before submission, while increasing volume and complexity of AI-generated reports prolong maintainer triage, leaving more recent issues without definitive verdicts.

\begin{figure}[t]
\centering
\includegraphics[width=0.9\columnwidth]{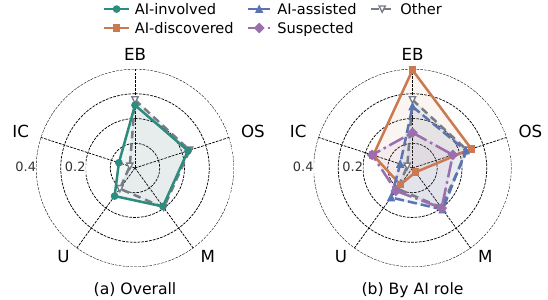}
\caption{Rejection-reason prevalence among invalid reports: (a) AI-involved versus Other; (b) AI roles versus Other. \textbf{EB}: Expected Behavior; \textbf{OS}: Out of Scope; \textbf{M}: Misuse; \textbf{U}: Unreproducible; \textbf{IC}: Incorrect Claims. Reasons are non-exclusive; the catch-all \emph{Others} reason is omitted.}
\label{fig:invalid-reasons}
\end{figure}

\noindent\textbf{Rejected AI-involved reports more often contain incorrect claims and unreproducible findings.}~
We examine six non-exclusive rejection reasons reported by maintainers: Misuse, Expected Behavior, Unreproducible, Incorrect Claims, Out-of-Scope, and Others. \autoref{fig:invalid-reasons} compares these reasons among invalid reports. Compared with other reports, rejected AI-involved reports are more likely to involve Incorrect Claims (+4.62 pp) and Unreproducible findings (+3.65 pp). The differences vary by AI role. Incorrect Claims are particularly common among AI-discovered (+14.44 pp) and suspected reports (+14.92 pp), while AI-assisted reports show a larger difference in Unreproducible findings (+4.35 pp). These patterns may reflect different challenges in AI-assisted bug reporting: AI-discovered reports often begin with code-level hypotheses requiring verification, whereas AI-assisted reports may struggle to reproduce failures encountered in practice.


\subsection{Factors Associated with Validity}
\label{subsec:validity-factors}
To identify which uses of AI and report attributes are associated with higher validity (Section~\ref{subsec:validity}), we estimate two Firth logistic regression models on 68,773 reports with definitive verdicts. The main model includes discovery method, six AI-task indicators, and report quality. The second replaces quality with seven missing-element indicators to avoid overlapping measures. Both models control for AI role, project, and quarter, with Operational Usage and Medium quality as reference categories. We use project-clustered standard errors and Wald tests, applying Benjamini--Hochberg correction to factor-group and coefficient-level tests. \autoref{tab:validity-factors} reports adjusted odds ratios (ORs), 95\% confidence intervals, and corrected $q$-values.

\begin{table}[t]
\centering
\caption{Factors associated with maintainer-confirmed validity.}
\label{tab:validity-factors}
\renewcommand{\arraystretch}{1.08}
\resizebox{0.9\columnwidth}{!}{%
\begin{tabular}{@{}lrrc@{}}
\toprule
\textbf{Factor} & \textbf{Valid (\%)} & \textbf{OR [95\% CI]} & \textbf{$q$} \\
\midrule
\multicolumn{4}{@{}l}{\textit{Main Model: Discovery Method ($q<0.001$)}} \\
\quad Operational Usage (ref.)
& 45.43\% & 1 & -- \\
\quad Source Code Review
& 79.13\% & \textbf{2.82} [2.26, 3.52] & $<0.001$ \\
\quad Fuzzing
& 72.65\% & \textbf{2.78} [1.43, 5.42] & 0.005 \\
\quad Testing
& 71.49\% & \textbf{2.25} [1.93, 2.63] & $<0.001$ \\
\quad Dynamic Analysis
& 65.55\% & \textbf{1.98} [1.18, 3.34] & 0.019 \\
\quad Build / Installation
& 53.45\% & 1.19 [1.01, 1.41] & 0.052 \\
\quad Static Analysis
& 51.93\% & 1.09 [0.69, 1.72] & 0.786 \\
\midrule
\multicolumn{4}{@{}l}{\textit{Main Model: AI Tasks ($q=0.021$)}} \\
\quad Reproduction and Testing
& 81.20\% & \textbf{2.25} [1.16, 4.38] & 0.026 \\
\quad Problem Diagnosis
& 65.92\% & \textbf{1.42} [1.07, 1.87] & 0.024 \\
\quad Fix Recommendation
& 64.58\% & 1.36 [1.02, 1.82] & 0.052 \\
\quad Report Writing
& 63.91\% & 1.05 [0.78, 1.41] & 0.786 \\
\quad Information Retrieval
& 56.80\% & 0.93 [0.72, 1.20] & 0.702 \\
\quad Debugging Support
& 51.02\% & 0.66 [0.39, 1.12] & 0.163 \\
\midrule
\multicolumn{4}{@{}l}{\textit{Main Model: Quality ($q<0.001$)}} \\
\quad Medium (ref.)
& 48.87\% & 1 & -- \\
\quad High
& 60.33\% & \textbf{1.59} [1.43, 1.76] & $<0.001$ \\
\quad Low
& 38.14\% & \underline{0.56} [0.51, 0.63] & $<0.001$ \\
\midrule
\multicolumn{4}{@{}l}{\textit{Missing-Element Model: Missing Elements ($q<0.001$)}} \\
\quad Component
& 36.14\% & \underline{0.67} [0.60, 0.74] & $<0.001$ \\
\quad Trigger
& 39.07\% & \underline{0.79} [0.72, 0.86] & $<0.001$ \\
\quad Reproduction Steps
& 44.61\% & \underline{0.83} [0.77, 0.90] & $<0.001$ \\
\quad Observed Behavior
& 40.71\% & \underline{0.83} [0.76, 0.91] & $<0.001$ \\
\quad Expected Behavior
& 44.93\% & \underline{0.80} [0.71, 0.89] & $<0.001$ \\
\quad Summary
& 41.91\% & 1.02 [0.91, 1.14] & 0.786 \\
\quad Environment
& 54.88\% & 1.01 [0.94, 1.07] & 0.838 \\
\bottomrule
\end{tabular}%
}

\parbox{\columnwidth}{\footnotesize
\vspace{1mm}
\textit{Note.} Bold ORs mark significant positive associations and underlined ORs negative associations at BH-adjusted $q<0.05$. CIs are unadjusted. Valid (\%) is the unadjusted valid rate among decided reports with each attribute.
}
\end{table}

\noindent\textbf{Systematic discovery methods are associated with higher validity.}~
Compared with Operational Usage, reports from Source Code Review, Fuzzing, Testing, and Dynamic Analysis have significantly higher validity odds, showing adjusted ORs of 2.82, 2.78, 2.25, and 1.98 respectively at $q<0.05$. Conversely, Build/Installation and Static Analysis show no significant associations. This suggests systematic methods yield concrete, verifiable test cases, while ad-hoc observations or unverified warnings are harder to confirm.

\noindent\textbf{AI-assisted reproduction and diagnosis are associated with higher validity.}~
AI tasks are jointly associated with validity ($q=0.021$). Reproduction and Testing (OR = 2.25, $q=0.026$) and Problem Diagnosis (OR = 1.42, $q=0.024$) show significant positive associations. Conversely, Fix Recommendation, Report Writing, Information Retrieval, and Debugging Support lack significant associations. Thus, AI proves more useful for reproducing failures and investigating causes than merely generating descriptions or fixes, helping translate potential bugs into verifiable claims.

\noindent\textbf{Higher quality is associated with higher validity.}~
Report quality is significantly associated with validity ($q<0.001$). Compared to medium-quality reports, high-quality ones have a higher valid verdict odds ratio of 1.59, whereas low-quality reports drop to an OR of 0.56. This pattern reflects lower triage friction: well-structured reports reduce the effort required for maintainers to understand and verify an issue, whereas incomplete descriptions risk early dismissal.

\noindent\textbf{Missing component and reproduction information is associated with lower validity.}~
Missing information elements are jointly associated with validity ($q<0.001$). Missing component information shows the strongest negative association, with an OR of 0.67. Missing triggering conditions, expected behavior, reproduction steps, and observed behavior are associated with lower odds of validity, with ORs of 0.79, 0.80, 0.83, and 0.83, respectively. All five associations remain significant after correction ($q<0.001$). In contrast, missing summary and environment information show no significant associations. These results highlight the importance of identifying the affected component and providing clear conditions for reproducing the problem.

\begin{tcolorbox}[
    enhanced,
    colback=black!6,
    colframe=black,
    boxrule=0.5pt,
    arc=3mm,
    left=5pt,
    right=5pt,
    top=4pt,
    bottom=4pt,
    before skip=8pt,
    after skip=8pt
]
\noindent
\faLightbulb[regular]\quad
\textbf{Key Takeaways (RQ2):}~
AI-involved reports are more complete and show rising validity among decided cases over time, yet overall validity does not surpass other reports as uncertain verdicts grow and rejections more often involve incorrect claims or unreproducible steps. Higher validity is linked to systematic discovery, AI-assisted reproduction and diagnosis, and complete localization and reproduction details.
\end{tcolorbox}

\begin{figure}[t]
    \centering
    \subfloat[Overall.\label{fig:closure-overall}]{%
        \includegraphics[width=0.49\columnwidth]{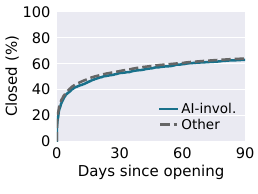}}%
    \hfill
    \subfloat[By AI role.\label{fig:closure-roles}]{%
        \includegraphics[width=0.49\columnwidth]{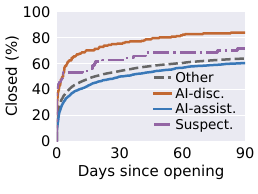}}
    \caption{Observed cumulative closure rates within 90 days.}
    \label{fig:closure}
\end{figure}
\section{How Does AI Influence the Handling of Bug Reports?}
\label{sec:maintainers}

\begin{figure}[t]
    \centering
    \subfloat[Overall.\label{fig:replies-overall}]{%
        \includegraphics[width=0.49\columnwidth]{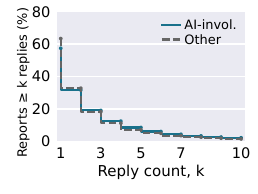}}%
    \hfill
    \subfloat[By AI role.\label{fig:replies-roles}]{%
        \includegraphics[width=0.49\columnwidth]{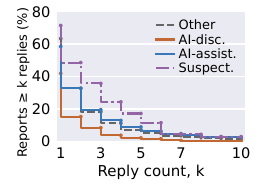}}
    \caption{Share of reports with at least $k$ maintainer replies within 30 days.}
    \label{fig:replies}
\end{figure}

As shown in previous sections, AI-involved reports provide more complete information (Section~\ref{subsec:information-completeness}) but also pose distinct verification challenges (Section~\ref{subsec:validity}). We examine how this dual nature reshapes maintainer workflows through a sequence of cascading effects: (1) \textit{Maintainer attention}, measured by closure rates and replies to gauge initial engagement; (2) \textit{Distribution of maintenance effort}, examining how this attention translates into actions across triage, validation, and resolution; (3) \textit{Material evaluation}, analyzing how maintainers scrutinize specific AI-generated artifacts during substantive validation; and (4) \textit{Maintainer-side AI use}, exploring how this growing verification burden motivates maintainers to integrate AI tools into their own investigative practices. Together, these analyses characterize how AI shifts the focus of maintenance work.

\subsection{Maintainer Attention}
\label{subsec:interactions}

We examine how maintainers allocate initial attention to AI-involved reports through closure rates and reply counts. We measure closure within 90 days and count maintainer replies within 30 days, excluding bots. On GitHub, we identify maintainers by project roles; for Linux Bugzilla, we use non-reporter comments as a proxy. Both analyses include 2,117 AI-involved and 109,326 other reports with at least 90 days of observation. We compare groups within project-quarters and apply the statistical procedures in Section~\ref{subsec:statistical-analysis}.

\noindent\textbf{Overall closure rates show no significant difference, but AI-assisted reports remain open longer.}
As shown in Figure~\ref{fig:closure-overall}, closure rates are similar between AI-involved and other reports ($q\geq0.272$). However, the patterns differ across AI roles. AI-assisted reports have significantly lower closure rates within 30 and 90 days, with adjusted differences of $-3.43$ and $-3.27$ percentage points. AI-discovered reports have higher observed closure rates, but the adjusted differences are not significant. Rather than indicating maintainer neglect, this delayed closure aligns with our subsequent finding in Section~\ref{subsec:actions} that AI-assisted reports trigger more time-consuming validation activities.

\noindent\textbf{AI-discovered reports struggle to attract maintainer attention and often close without definitive verdicts.}
Figure~\ref{fig:replies-overall} shows that AI-involved reports are less likely to receive any reply, while the distributions become similar at higher reply counts. As shown in Figure~\ref{fig:replies-roles}, AI-discovered reports receive the fewest replies, whereas AI-assisted reports follow patterns similar to other reports. Suspected reports are more likely to involve extended discussions. These differences may reflect varying levels of maintainer engagement and opportunities for follow-up investigation.

The lack of engagement on AI-discovered reports is closely tied to their origin. As noted in Section~\ref{subsec:discovery-sources}, AI-discovered reports often begin as code-level hypotheses from source code review rather than failures observed during software use. Consequently, among the 160 AI-discovered reports closed within 90 days, 96 remain uncertain, including 71 without any follow-up comments. Lacking clear real-world impact or an active reporter to provide operational context, maintainers have minimal incentive to investigate these automated findings, leading many to be ignored or dismissed without a definitive verdict.

\subsection{Maintainer Actions}
\label{subsec:actions}

To understand whether AI creates an extra burden or facilitates faster processing, we further examine maintainer actions when handling bug reports. We classify maintainer replies into six actions across three stages. Triage covers Information Request and Duplicate Checking; Validation covers Reproduction and Testing and Code and API Investigation; Resolution covers Fix Recommendation and Patch Review and Testing. Detailed in Appendix~\ref{sec:explore}, our exploratory analysis reveals that higher information completeness intrinsically shifts maintainer effort, reducing basic information requests while increasing substantive validation and resolution activities. We count actions per report within 30 days of submission, using the same cohort as Section~\ref{subsec:interactions}. We exclude Linux Bugzilla because it lacks explicit maintainer labels, while official maintainer lists may be outdated. We compare groups within project-quarters using the procedures in Section~\ref{subsec:statistical-analysis}. Results appear in \autoref{tab:actions}.

\begin{table*}[t]
\centering
\caption{Maintainer actions per report within 30 days, by AI role.}
\label{tab:actions}

\scriptsize
\setlength{\tabcolsep}{3pt}
\begin{tabular*}{0.95\textwidth}{@{\extracolsep{\fill}}ll ccc ccc ccc c@{}}
\toprule
\multirow{2}{*}{\textbf{AI Role}}
& \multirow{2}{*}{\textbf{Metric}}
& \multicolumn{3}{c}{\textbf{Triage}}
& \multicolumn{3}{c}{\textbf{Validation}}
& \multicolumn{3}{c}{\textbf{Resolution}}
& \multirow{2}{*}{\textbf{Total}} \\
\cmidrule(lr){3-5}
\cmidrule(lr){6-8}
\cmidrule(lr){9-11}
& & \textbf{Info.} & \textbf{Dup.} & \textbf{Total}
& \textbf{Repro.} & \textbf{Code} & \textbf{Total}
& \textbf{Fix} & \textbf{Patch} & \textbf{Total}
& \\
\midrule

\textbf{Other} & Mean
& 0.28 & 0.13 & 0.41
& 0.30 & 0.70 & 1.01
& 0.53 & 0.05 & 0.58 & 2.00 \\
\midrule

\multirow{2}{*}{\textbf{AI-disc.}}
& Mean
& 0.06 & 0.06 & 0.12
& 0.10 & 0.37 & 0.47
& 0.29 & 0.05 & 0.34 & 0.93 \\
& Adj. $\Delta$
& \textbf{\textminus0.12}\textsuperscript{*}
& \textminus0.03
& \textbf{\textminus0.15}\textsuperscript{*}
& \textbf{\textminus0.17}\textsuperscript{*}
& \textminus0.27
& \textbf{\textminus0.44}\textsuperscript{*}
& \textbf{\textminus0.30}\textsuperscript{*}
& \textminus0.03
& \textbf{\textminus0.34}\textsuperscript{*}
& \textbf{\textminus0.93}\textsuperscript{*} \\

\addlinespace[3pt]
\multirow{2}{*}{\textbf{AI-assist.}}
& Mean
& 0.32 & 0.11 & 0.44
& 0.34 & 0.73 & 1.06
& 0.61 & 0.07 & 0.68 & 2.18 \\
& Adj. $\Delta$
& \textbf{+0.10}\textsuperscript{*}
& +0.01
& \textbf{+0.11}\textsuperscript{*}
& \textbf{+0.08}\textsuperscript{*}
& \textbf{+0.15}\textsuperscript{*}
& \textbf{+0.24}\textsuperscript{*}
& +0.09
& +0.02
& +0.11
& \textbf{+0.46}\textsuperscript{*} \\

\addlinespace[3pt]
\multirow{2}{*}{\textbf{Suspect.}}
& Mean
& 0.28 & 0.09 & 0.37
& 0.28 & 0.91 & 1.19
& 0.57 & 0.07 & 0.65 & 2.21 \\
& Adj. $\Delta$
& \textminus0.06 & \textminus0.06 & \textminus0.12
& \textminus0.09 & +0.02 & \textminus0.06
& \textminus0.13 & +0.02 & \textminus0.11
& \textminus0.29 \\

\midrule
\multirow{2}{*}{\textbf{AI-invol.}}
& Mean
& 0.30 & 0.11 & 0.41
& 0.31 & 0.70 & 1.01
& 0.58 & 0.07 & 0.65 & 2.07 \\
& Adj. $\Delta$
& +0.07 & 0.00 & +0.07
& +0.05 & +0.11 & +0.16
& +0.05 & +0.02 & +0.07 & +0.30 \\
\bottomrule
\end{tabular*}

\vspace{1mm}
\parbox{0.95\textwidth}{\footnotesize
\textit{Note.} Actions are non-exclusive.
Bold values with \textsuperscript{*} indicate
Benjamini--Hochberg adjusted q-values below 0.05.
Confidence intervals and exact q-values are reported in
\autoref{tab:actions-ci}. \textbf{Info.}: information request; \textbf{Dup.}: duplicate checking; \textbf{Repro.}: reproduction and testing; \textbf{Code}: code and API investigation; \textbf{Fix}: fix recommendation; \textbf{Patch}: patch review and testing; \textbf{Adj}.: project-quarter adjusted difference ($\Delta$).
}
\end{table*}

\noindent\textbf{AI-involved reports show no significant overall increase in maintainer actions.}
As shown in \autoref{tab:actions}, AI-involved reports receive 0.30 more maintainer actions per report after adjusting for project and time, but the difference is not statistically significant across stages. However, this aggregate finding masks two contrasting dynamics: AI-assisted reports drive deeper maintainer engagement and effort, whereas AI-discovered reports experience a severe drop in attention.

\noindent\textbf{AI-assisted reports effectively engage maintainers, driving deeper validation and diagnostic discussions.}~
Compared with other reports from the same project and quarter, AI-assisted reports receive 0.46 additional actions per report ($q=0.018$). This increase spans both Validation (+0.24, $q=0.018$) and Triage (+0.11, $q=0.018$). Because AI-assisted reports generally feature higher information completeness (Section~\ref{subsec:information-completeness}), they provide the necessary structural details to immediately trigger substantive technical investigation, leading to significant increases in Code and API Investigation (+0.15) and Reproduction and Testing (+0.08). 

Paradoxically, this completeness coincides with significantly more Information Requests (+0.10, $q=0.018$). To mitigate the potential flaws of AI-generated content (Section~\ref{subsec:maintainer-material}), maintainers must ask follow-up questions to disentangle actual observations from AI hypotheses. Ultimately, while AI-assisted reports trigger deep technical validation, they still require extra effort to filter out generated noise.

\noindent\textbf{AI-discovered reports receive fewer actions, reflecting both reduced burden and limited engagement.}~
While the high structural completeness of AI-discovered reports naturally reduces the need for basic Triage clarification (-0.15, $q=0.018$), they also experience massive drops in substantive Validation (-0.44, $q=0.018$) and Resolution (-0.34, $q=0.018$) activities. This sweeping decline in later-stage effort, combined with the fact that fewer than 20\% receive even a single reply (Section~\ref{subsec:interactions}), indicates a severe lack of maintainer attention rather than easily resolved bugs. Even though AI agents generate well-structured defect hypotheses, their lack of real-world user impact or an active reporter to provide operational context gives maintainers minimal incentive to begin validation, leaving many of these automated findings practically ignored.

\subsection{Material Evaluation}
\label{subsec:maintainer-material}
While AI-involved reports share comparable validity rates (Section~\ref{subsec:validity}), confirming a bug exists does not equate to endorsing its accompanying diagnosis or fix. To unpack this distinction, we examine how maintainers evaluate three types of materials: \emph{Reproducers}, \emph{Technical claims}, and \emph{Proposed fixes}. Indeed, even within maintainer-confirmed \emph{valid} reports, 32.65\% of technical claims and 33.76\% of proposed fixes are ultimately rejected, detailed in Appendix~\ref{sec:explore}. This indicates that while AI effectively flags genuine defects, its detailed artifacts often require substantial human correction.

To compare how explicitly AI-generated materials are handled, we identified 76,028 materials across our dataset, including 265 explicitly AI-generated ones (34 reproducers, 148 claims, 83 fixes). An LLM classified maintainer responses as \emph{Use}, \emph{Request supplementation}, or \emph{Reject}. Given the small AI-generated sample, we applied a quarter-stratified exact conditional test, reporting conditional odds ratios and exact 95\% confidence intervals with Benjamini--Hochberg correction across nine comparisons (\autoref{tab:material-actions}).

\begin{table}[t]
\centering
\caption{Maintainer actions on AI-generated versus other report materials.}
\label{tab:material-actions}
\footnotesize
\renewcommand{\arraystretch}{1.12}
\resizebox{0.9\columnwidth}{!}{%
\begin{tabular}{@{}l c c c c@{}}
\toprule
\textbf{Material / Action} & \textbf{AI-Gen} & \textbf{Other} & \textbf{OR [95\% CI]} & \textbf{$q$} \\
\midrule
\multicolumn{5}{@{}l}{\textit{Reproducer}} \\
\quad Reject & 23.53\% & 5.14\% & \textbf{8.16 [3.15, 18.86]}\textsuperscript{*} & \textbf{\boldmath$<$.001} \\
\quad Req.\ supp. & 38.24\% & 58.17\% & 0.52 [0.24, 1.09] & 0.147 \\
\quad Use & 38.24\% & 27.47\% & 1.31 [0.60, 2.76] & 0.650 \\
\midrule
\multicolumn{5}{@{}l}{\textit{Technical claim}} \\
\quad Reject & 59.46\% & 54.72\% & \textbf{1.57 [1.11, 2.23]}\textsuperscript{*} & \textbf{0.034} \\
\quad Req.\ supp. & 8.78\% & 12.49\% & 0.79 [0.41, 1.40] & 0.650 \\
\quad Use & 6.76\% & 3.20\% & 0.92 [0.42, 1.78] & 1.000 \\
\midrule
\multicolumn{5}{@{}l}{\textit{Proposed fix}} \\
\quad Reject & 54.22\% & 49.76\% & 1.55 [0.97, 2.48] & 0.123 \\
\quad Req.\ supp. & 13.25\% & 16.63\% & 0.99 [0.47, 1.92] & 1.000 \\
\quad Use & 16.87\% & 20.73\% & \textbf{0.49 [0.25, 0.90]}\textsuperscript{*} & \textbf{0.0498} \\
\bottomrule
\end{tabular}}

\parbox{0.9\linewidth}{
\vspace{1mm}
\footnotesize
\textit{Notes.} \textsuperscript{*} marks $q < 0.05$ after Benjamini--Hochberg correction. \textbf{AI-Gen}: explicitly AI-generated materials; \textbf{Req. supp.}: request for supplementation. ORs compare materials within the same quarter.
}
\end{table}

\noindent\textbf{AI-generated materials are more frequently rejected, reinforcing the need for human validation.}
As shown in \autoref{tab:material-actions}, significant differences appear mainly in how maintainers reject technical materials. Among reproducers, 23.53\% of AI-generated materials are rejected, compared with 5.14\% of other materials (OR = 8.16, $q<0.001$). In addition, AI-generated technical claims are also challenged more frequently, and proposed fixes are less often used by maintainers. These frequent rejections often occur because human reporters provide insufficient execution context, forcing AI to rely on ungrounded assumptions that fail during testing. This context gap explains the increased validation time required for AI-assisted reports (Section~\ref{subsec:actions}). Ultimately, while AI provides structural completeness, generating reliable technical artifacts fundamentally demands active human verification.

\subsection{Maintainers Using AI}
\label{subsec:maintainer-ai}

Because AI-assisted reports trigger more deep technical validation (Section~\ref{subsec:actions}) and AI-generated claims require meticulous verification (Section~\ref{subsec:maintainer-material}), maintainers themselves are increasingly adopting AI to help shoulder this investigative workload. We analyze replies from GitHub project owners, members, collaborators, and contributors, excluding the original reporters. A reply is classified as \emph{AI-involved} if it is written by an AI bot or the maintainer explicitly acknowledges using AI. We distinguish between AI bots and maintainers using AI, and exclude replies that merely discuss AI-generated content from reporters.

\begin{figure}[t]
\centering
\includegraphics[width=\columnwidth]{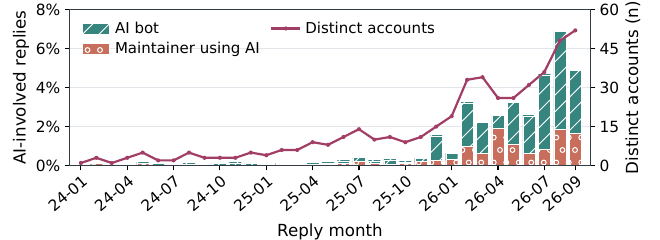}
\caption{AI-involved maintainer replies by author type and distinct accounts, grouped by reply month.}
\label{fig:maintainer-ai-trend}
\end{figure}

\noindent\textbf{AI adoption in maintenance is growing but remains concentrated.}~
Among 224,979 maintainer replies, 2,291 involve AI across 1,573 bug reports. As shown in \autoref{fig:maintainer-ai-trend}, the share of AI-involved replies increased from 0.07\% in 2024 to 0.31\% in 2025 and 3.40\% in the first nine months of 2026, reaching 6.88\% in August 2026, while overall reply volume remained around 6,000 per month. AI adoption expanded from 19 to 184 accounts and from 14 to 69 projects between 2024 and 2026, indicating participation beyond early adopters. AI bots account for 65.4\% of AI-involved replies, compared with 34.6\% from human maintainers using AI, with both groups growing substantially. However, adoption remains uneven, with five projects accounting for 63.6\% of AI-involved replies in 2026, suggesting that project-specific bot integration plays an important role.

\noindent\textbf{Maintainers use AI mainly for code analysis and fix recommendations.}
Validation accounts for 55.92\% of AI-involved maintainer actions, followed by Resolution at 31.98\% and Triage at 12.10\%. Code and API Investigation is the most common action at 40.10\%, followed by Fix Recommendation at 28.20\% and Reproduction and Testing at 15.82\%. Similar patterns appear for AI bots and maintainers using AI. AI contributes to 1.6\%--2.1\% of most maintainer action types, but only 0.65\% of Information Requests. This may reflect where AI can operate independently. Code analysis and fix suggestions can build on source code and report content, while information requests depend on details only reporters can provide. AI therefore appears more useful for examining available evidence than for resolving missing information through communication.

\begin{figure}[t]
\centering
\includegraphics[width=0.75\columnwidth]{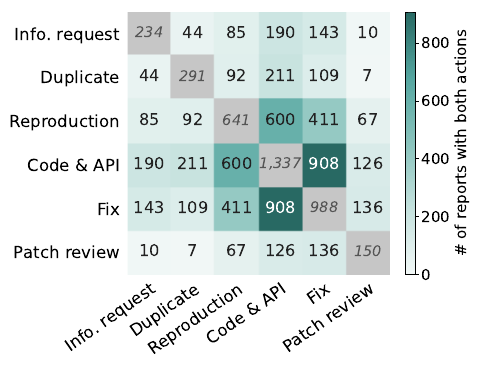}
\vspace{-2mm}
\caption{Action co-occurrence in AI-involved maintainer replies. Cells count reports with each action on the diagonal and both actions elsewhere.}
\vspace{-2mm}
\label{fig:maintainer-ai-cooccur}
\end{figure}

\noindent\textbf{Maintainers often use AI for multiple related tasks.}
Among the 1,534 reports containing AI-involved maintainer actions, 82.0\% involve at least two action types. As shown in \autoref{fig:maintainer-ai-cooccur}, Code and API Investigation frequently co-occurs with Fix Recommendation in 908 reports and Reproduction and Testing in 600 reports. Reproduction and Testing also co-occurs with Fix Recommendation in 411 reports. These combinations suggest that maintainers use AI not just to answer isolated questions, but to connect different parts of bug handling. For example, identifying a possible cause can lead directly to generating a reproducer or suggesting a fix.

\noindent\textbf{Maintainers use AI more often when handling AI-assisted reports.}
As shown in \autoref{tab:maintainer-ai-compare}, 4.48\% of AI-involved reports receive AI-involved maintainer replies, compared with 1.26\% of other reports. However, the adjusted difference is not significant ($q=0.131$). The patterns differ between AI bots and maintainers using AI. Bot replies show no significant difference across report types, suggesting that they largely follow project-level automation workflows. In contrast, replies from maintainers using AI are more common on AI-involved reports, with an adjusted difference of +1.33 percentage points ($q=0.003$). This increase is driven by AI-assisted reports (+1.58 points, $q=0.005$), particularly during Validation (+1.64 points, $q=0.009$). This pattern may reflect how AI-assisted reports shape subsequent maintenance work. These reports often contain AI-supported diagnoses and fix suggestions, which maintainers may use AI to examine, verify, or refine. As AI-assisted reports also receive more validation actions, maintainers may be turning to AI to help assess claims that were themselves developed with AI. This suggests that AI use on the reporter side may encourage further AI use during maintenance, creating a workflow in which both sides rely on AI to investigate the same bug.

\begin{table}[t]
\centering
\caption{AI-involved maintainer replies and associated actions by report AI role.}
\label{tab:maintainer-ai-compare}
\renewcommand{\arraystretch}{1.15}
\setlength{\tabcolsep}{4pt}
\resizebox{\columnwidth}{!}{%
\begin{tabular}{@{}ll ccc ccc@{}}
\toprule
\multirow{2}{*}{\textbf{AI Role}}
& \multirow{2}{*}{\textbf{Metric}}
& \multicolumn{3}{c}{\textbf{AI-Involved Replies}}
& \multicolumn{3}{c}{\textbf{Actions in AI-Involved Replies}} \\
\cmidrule(lr){3-5}
\cmidrule(lr){6-8}
& & \textbf{Any} & \textbf{AI Bot} & \textbf{Maint. AI}
& \textbf{Triage} & \textbf{Validation} & \textbf{Resolution} \\
\midrule
\textbf{Other}
& Value
& 1.26 & 0.86 & 0.47
& 0.39 & 1.10 & 0.79 \\
\midrule

\multirow{2}{*}{\textbf{AI-disc.}}
& Value
& 2.28 & 1.27 & 1.01
& 0.00 & 1.77 & 1.77 \\
& Adj. $\Delta$
& $-2.09$ & $-2.13$ & $-0.26$
& $-0.76$ & $-2.35$ & $-1.83$ \\

\addlinespace[3pt]
\multirow{2}{*}{\textbf{AI-assist.}}
& Value
& 4.89 & 2.42 & 3.04
& 1.15 & 4.44 & 3.41 \\
& Adj. $\Delta$
& $\mathbf{+1.56}^{*}$ & $+0.18$ & $\mathbf{+1.58}^{*}$
& $+0.33$ & $\mathbf{+1.64}^{*}$ & $+1.05$ \\

\addlinespace[3pt]
\multirow{2}{*}{\textbf{Suspect.}}
& Value
& 3.19 & 1.06 & 2.13
& 1.06 & 2.13 & 2.13 \\
& Adj. $\Delta$
& $+2.20$ & $+0.69$ & $+1.46$
& $+0.85$ & $+1.44$ & $+1.44$ \\

\midrule
\multirow{2}{*}{\textbf{AI-invol.}}
& Value
& 4.48 & 2.22 & 2.74
& 0.99 & 4.00 & 3.15 \\
& Adj. $\Delta$
& $+1.09$ & $-0.12$ & $\mathbf{+1.33}^{*}$
& $+0.20$ & $+1.09$ & $+0.67$ \\
\bottomrule
\end{tabular}%
}

\parbox{\columnwidth}{\footnotesize
\vspace{1mm}
\textit{Note.} Bold differences with $^{*}$ indicate $q<0.05$ after Benjamini--Hochberg correction. Values are percentages of reports in GitHub projects. Adjusted $\Delta$ compares each group with Other within project-quarters.
}
\end{table}

\begin{tcolorbox}[
    enhanced,
    colback=black!6,
    colframe=black,
    boxrule=0.5pt,
    arc=3mm,
    left=5pt,
    right=5pt,
    top=4pt,
    bottom=4pt,
    before skip=8pt,
    after skip=8pt
]
\noindent
\faLightbulb[regular]\quad
\noindent\textbf{Key Takeaways (RQ3).}~
Maintainer effort diverges by AI role: AI-assisted reports drive deeper investigation, whereas AI-discovered reports receive less attention. Moreover, even when the reported bugs are valid, generated materials could be wrong, highlighting the need for human verification. Meanwhile, AI use in bug reporting also encourages maintainers to adopt AI in their own maintenance workflows.
\end{tcolorbox}
\section{Implications and Recommendations}
\label{sec:implications}

Our findings show that AI-involved reports are becoming more complete and reliable, with AI-assisted reproduction and diagnosis associated with higher validity. Meanwhile, maintainers increasingly use AI for investigation and fixes, but devote more effort to AI-assisted reports than to AI-discovered reports. These findings motivate practical recommendations for reporters and maintainers.

\subsection{Recommendations for Reporters}
\label{subsec:rec-reporters}

We propose principles for improving AI-involved bug reporting:
\begin{enumerate}
    \item \textbf{Reproducibility.} Provide sufficient details for maintainers to locate and reproduce the problem.
    \item \textbf{Verification.} Independently test generated reproducers and verify claims before submission.
    \item \textbf{Accountability.} Disclose AI involvement and ensure that a human remains responsible for report submission and follow-up.
\end{enumerate}

\noindent\textbf{Reproducibility.}~Our findings show that missing component and trigger information is associated with lower odds of a valid verdict (\autoref{tab:validity-factors}). Reporters should therefore name the affected command, API, or component and specify the tested release or commit. They should also provide the smallest input that reliably triggers the failure, along with the exact command to run it. The report should clearly state the expected behavior and include the actual output. For crashes or exceptions, include the error message and stack trace, highlighting the first relevant project frame to help maintainers identify where to start.

\noindent\textbf{Verification.}~Maintainers rejected 23.53\% of AI-generated reproducers, compared with 5.14\% of others, and more frequently challenged AI-generated technical claims (\autoref{tab:material-actions}). To avoid these problems, reporters should run generated tests on the affected version, confirm that they trigger the reported failure, and provide the inputs, commands, and observed outputs. Technical claims about API behavior, execution paths, root causes, and potential impacts should be checked against source code, documentation, and runtime results. For example, AI may claim that an API accepts invalid inputs while overlooking its documented contract, or infer a security vulnerability from a crash without sufficient support. Claims that remain unverified should be labeled as hypotheses and distinguished from confirmed findings.

\noindent\textbf{Accountability.}~Reports of bugs discovered by AI received fewer maintainer replies, and 96 of the 160 reports closed within 90 days had no clear verdict (Section~\ref{subsec:interactions}). One possible explanation is the lack of a human contact responsible for follow-up. When reporting a bug initially encountered by a human, reporters should describe the original failure and explain how AI contributed, such as through diagnosis or test generation. For bugs first identified by an agent, the finding should remain a draft until a person verifies the main claim and approves submission. That person should also answer maintainer questions, rerun tests, and update the report as needed. In both cases, reporters should identify the model or agent if known, describe its contribution, and distinguish independently verified findings from claims requiring further confirmation.

\subsection{Recommendations for Maintainers}
\label{subsec:rec-maintainers}

We propose principles for handling AI-involved reports:
\begin{enumerate}
    \item \textbf{Adaptation.} Recognize the growing impact of AI on bug reporting and establish clear policies for AI-involved submissions.
    \item \textbf{Standardization.} Establish structured reporting formats that separate observed failures from AI-generated analyses and proposed fixes.
    \item \textbf{Prevalidation.} Identify common errors and unsupported claims before deeper investigation.
    \item \textbf{Traceability.} Preserve bot investigation results for verification and subsequent review.
\end{enumerate}

\noindent\textbf{Adaptation.}~
The share of AI-involved reports rose from 0.7\% in 2024 to 8.3\% in 2026--Q3, while completeness improved over time (Figures~\ref{fig:ai-growth} and~\ref{fig:report-quality-timeline}). In response, maintainers should establish clear and consistent reporting guidelines that evaluate reports based on technical content rather than AI involvement alone. These guidelines should require reporters to disclose how AI was used and which findings were independently verified. For automated submissions, require a human contact responsible for answering questions, rerunning tests, and updating the report.

\noindent\textbf{Standardization.}~
AI-assisted reports received more validation actions than other reports after adjustment (Section~\ref{subsec:actions}). To facilitate initial triage, maintainers should require reporters to present the \textit{original observed failure} before AI-generated diagnoses or proposed fixes. This includes the affected version, triggering input or command, and relevant output. Proposed causes and fixes should appear in a separate section. For lengthy submissions, ask reporters to extract relevant error messages or log lines rather than leaving them embedded in AI-generated explanations.

\noindent\textbf{Prevalidation.}~
Among invalid AI-involved reports, expected behavior and out-of-scope issues accounted for 25.38\% and 22.54\%, respectively, as non-exclusive rejection reasons (\autoref{fig:invalid-reasons}). Before investigation, maintainers should check whether the reported behavior constitutes a bug under supported usage. For API failures, examine whether the behavior violates the documented contract. For findings from source code inspection, determine whether the suspected execution path is reachable. For example, if AI identifies a potential null dereference, check whether a supported API call can pass null to the affected branch. If confirmed, proceed with investigation; otherwise, document the unverified condition and request further information.

\noindent\textbf{Traceability.}~
Code and API investigation accounted for 40.10\% of AI-involved maintainer actions (Section~\ref{subsec:maintainer-ai}). Maintainers can delegate tasks such as tracing callers, inspecting input checks, or testing reproducers across software versions to bots. However, bots should record the \textit{inspected commits, relevant code locations, executed commands, test results, and unresolved questions} so that subsequent maintainers can verify and reuse their findings. For example, when a bot identifies a potential regression, maintainers should verify that the same input was tested across versions and that the reported failure occurred before using the result to guide further investigation.

\subsection{Applying the Reporting Principles}
\label{subsec:reporting-practice}

We applied the three reporting principles through an internal review workflow for submitting 53 reports of AI-discovered bugs through fuzzing. For \emph{reproducibility}, each report followed a common template specifying the tested version, affected component, minimized reproducer, build and execution commands, and resulting stack trace. For \emph{verification}, submitters reran reproducers against the latest upstream version, while internal reviewers examined API contracts, call sequences, and memory ownership to identify potential harness misuse before submission. For \emph{accountability}, reports disclosed AI involvement and human verification, with submitters responsible for responding to maintainer feedback and tracking resolution. As of writing, 69.81\% of the reports had received valid maintainer verdicts, reaching 94.87\% among reports with definitive verdicts. Most remaining reports lacked definitive verdicts. Of the two rejected reports, one concerned a robustness issue that maintainers did not consider worth fixing, while the other involved failures reproducible only with sanitizers enabled.
\section{Conclusion}
\label{sec:conclusion}
We conduct a large-scale study of AI involvement in 120,918 bug reports from 101 open-source projects, tracing its role from discovery and reporting through validation and resolution. Our findings show that AI-involved reports are becoming more complete and reliable over time, although greater completeness does not necessarily imply higher validity. AI-assisted reproduction and diagnosis are associated with higher validity, while generated claims still require verification. Maintainer practices differ by AI role: AI-assisted reports receive more validation actions, whereas AI-discovered reports receive less follow-up. Maintainers also increasingly use AI for code investigation and fix recommendations. These findings demonstrate AI's growing role in bug reporting and maintenance and the need for reproducibility, verification, and human accountability. Our code and data are available at \href{https://github.com/Norrrrrrr-lyn/AI-bug-report}{\textcolor{blue}{\nolinkurl{https://github.com/Norrrrrrr-lyn/AI-bug-report}}}.
\bibliographystyle{IEEEtran}
\bibliography{references}

\appendices
\setcounter{dbltopnumber}{4}
\setcounter{topnumber}{3}
\renewcommand{\dbltopfraction}{0.9}
\renewcommand{\topfraction}{0.9}
\renewcommand{\textfraction}{0.1}
\section{Temporal Trend Analysis}
\label{app:temporal-analysis}
We estimate quarterly trends among AI-involved reports (2024Q1--2026Q3), controlling for project and AI role, with project-clustered standard errors and finite-sample correction. Binary outcomes use linear probability models. Two-sided Wald tests with BH correction cover six outcomes: high completeness, mean missing elements, valid, invalid, uncertain, and valid among decided, with separate correction families by specification and sample. Sensitivity analyses exclude 2026Q3 or suspected reports and vary model controls. Quarterly increases in high completeness (+5.41 pp, 95\% CI [4.40, 6.41]) and validity among decided (+4.89 pp, 95\% CI [3.70, 6.07]) remain significant across exclusions ($q<0.001$). \autoref{fig:report-quality-timeline} shows unadjusted rates with 5,000 project-bootstrap percentile intervals.

\section{Detailed Verdicts and Confidence Intervals}
\label{app:ci}

\noindent\textbf{Report Validity.}
\autoref{tab:validity} shows that AI-involved reports have a higher observed validity rate among decided reports (61.00\% vs. 51.28\%), but the adjusted difference is not significant. AI-discovered reports have a significantly lower adjusted valid rate ($-10.92$ pp, $q=0.033$) and the highest proportion of uncertain verdicts (62.37\%). No significant differences appear for AI-assisted or suspected reports.

\noindent\textbf{Maintainer Actions.}
\autoref{tab:actions-ci} confirms contrasting patterns across AI roles. AI-assisted reports receive significantly more maintainer actions (95\% CI [0.21, 0.85], $q=0.018$), particularly during validation. In contrast, AI-discovered reports receive fewer actions (95\% CI [$-1.54$, $-0.25$], $q=0.018$), with significant reductions across all three stages. The overall difference for AI-involved reports remains insignificant ($q=0.061$).

\begin{table}[t]
\centering
\caption{Bug report validity by AI role.}
\label{tab:validity}
\scriptsize
\setlength{\tabcolsep}{3pt}
\renewcommand{\arraystretch}{1.12}
\resizebox{0.85\columnwidth}{!}{%
\begin{tabular}{@{}llcccc@{}}
\toprule
\textbf{AI Role} & \textbf{Stat.}
& Valid & Invalid & Uncertain & Valid/Dec. \\
\midrule

\multirow{4}{*}{AI-disc.}
& Rate (\%) & 25.51 & 12.12 & 62.37 & 67.79 \\
& $\Delta$ (pp) & $-10.92^{*}$ & $-2.32$ & $+13.25$ & $-3.73$ \\
& 95\% CI & $[-16.8,-2.5]$ & $[-8.3,4.5]$
& $[1.2,22.9]$ & $[-13.3,6.3]$ \\
& $q$ & 0.033 & 0.819 & 0.102 & 0.819 \\
\midrule

\multirow{4}{*}{AI-assist.}
& Rate (\%) & 32.46 & 20.57 & 46.97 & 61.20 \\
& $\Delta$ (pp) & $+0.12$ & $-0.31$ & $+0.19$ & $+0.39$ \\
& 95\% CI & $[-2.3,3.1]$ & $[-2.3,1.4]$
& $[-2.1,2.0]$ & $[-3.1,4.7]$ \\
& $q$ & 0.931 & 0.931 & 0.931 & 0.931 \\
\midrule

\multirow{4}{*}{Suspect.}
& Rate (\%) & 22.92 & 36.46 & 40.62 & 38.60 \\
& $\Delta$ (pp) & $-7.48$ & $+6.27$ & $+1.21$ & $-9.37$ \\
& 95\% CI & $[-16.9,3.1]$ & $[-3.6,16.1]$
& $[-11.7,15.0]$ & $[-21.3,3.7]$ \\
& $q$ & 0.428 & 0.523 & 0.931 & 0.428 \\
\midrule

\multirow{4}{*}{\textbf{AI-invol.}}
& Rate (\%) & 31.23 & 19.97 & 48.80 & 61.00 \\
& $\Delta$ (pp) & $-1.58$ & $-0.37$ & $+1.95$ & $-0.37$ \\
& 95\% CI & $[-3.83,1.17]$ & $[-2.51,1.57]$
& $[-0.62,4.48]$ & $[-3.70,3.67]$ \\
& $q$ & 0.436 & 0.845 & 0.436 & 0.845 \\
\midrule

\textbf{Other}
& Rate (\%) & 29.24 & 27.78 & 42.98 & 51.28 \\
\bottomrule
\end{tabular}%
}
\vspace{1mm}

\end{table}

\begin{table*}[tbp]
\centering
\caption{Adjusted differences in maintainer actions: 95\% bootstrap CIs in actions/report and BH-corrected $q$-values.}
\label{tab:actions-ci}
\resizebox{\textwidth}{!}{%
\begin{tabular}{@{}ll ccc ccc ccc c@{}}
\toprule
\multirow{2}{*}{\textbf{AI Role}}
& \multirow{2}{*}{\textbf{Metric}}
& \multicolumn{3}{c}{\textbf{Triage}}
& \multicolumn{3}{c}{\textbf{Validation}}
& \multicolumn{3}{c}{\textbf{Resolution}}
& \multirow{2}{*}{\textbf{Total}} \\
\cmidrule(lr){3-5}
\cmidrule(lr){6-8}
\cmidrule(lr){9-11}
& & \textbf{Info.} & \textbf{Dup.} & \textbf{Total}
& \textbf{Repro.} & \textbf{Code} & \textbf{Total}
& \textbf{Fix} & \textbf{Patch} & \textbf{Total}
& \\
\midrule
\multirow{2}{*}{\textbf{AI-disc.}} & 95\% CI & [\textminus{}0.18, \textminus{}0.04] & [\textminus{}0.08, 0.02] & [\textminus{}0.24, \textminus{}0.05] & [\textminus{}0.27, \textminus{}0.06] & [\textminus{}0.53, 0.00] & [\textminus{}0.79, \textminus{}0.08] & [\textminus{}0.48, \textminus{}0.08] & [\textminus{}0.08, 0.03] & [\textminus{}0.53, \textminus{}0.08] & [\textminus{}1.54, \textminus{}0.25] \\
 & $q$-value & 0.018 & 0.322 & 0.018 & 0.018 & 0.090 & 0.040 & 0.018 & 0.322 & 0.018 & 0.018 \\
\midrule
\multirow{2}{*}{\textbf{AI-assist.}} & 95\% CI & [0.05, 0.18] & [\textminus{}0.01, 0.03] & [0.05, 0.20] & [0.04, 0.15] & [0.07, 0.29] & [0.11, 0.44] & [0.02, 0.20] & [0.00, 0.05] & [0.03, 0.25] & [0.21, 0.85] \\
 & $q$-value & 0.018 & 0.598 & 0.018 & 0.018 & 0.023 & 0.018 & 0.085 & 0.218 & 0.085 & 0.018 \\
\midrule
\multirow{2}{*}{\textbf{Suspect.}} & 95\% CI & [\textminus{}0.19, 0.12] & [\textminus{}0.13, 0.02] & [\textminus{}0.28, 0.10] & [\textminus{}0.25, 0.12] & [\textminus{}0.36, 0.49] & [\textminus{}0.57, 0.55] & [\textminus{}0.42, 0.31] & [\textminus{}0.06, 0.12] & [\textminus{}0.46, 0.36] & [\textminus{}1.18, 0.79] \\
 & $q$-value & 0.598 & 0.218 & 0.337 & 0.488 & 0.911 & 0.851 & 0.598 & 0.790 & 0.660 & 0.649 \\
\midrule
\multirow{2}{*}{\textbf{AI-invol.}} & 95\% CI & [0.03, 0.14] & [\textminus{}0.01, 0.02] & [0.03, 0.15] & [0.01, 0.11] & [0.03, 0.23] & [0.05, 0.33] & [\textminus{}0.02, 0.15] & [0.00, 0.04] & [\textminus{}0.01, 0.18] & [0.08, 0.63] \\
 & $q$-value & 0.061 & 0.951 & 0.061 & 0.061 & 0.061 & 0.061 & 0.259 & 0.247 & 0.247 & 0.061 \\
\bottomrule
\end{tabular}%
}
\end{table*}

\section{Maintainer Action and Material Usage}
\label{sec:explore}
In our exploratory analysis, information completeness significantly reshapes maintainer activity composition. Specifically, compared to low-completeness reports, high-completeness reports exhibit a notable reduction in basic information requests (an adjusted difference of $-0.145$ actions per report, 95\% CI $[-0.181, -0.108]$). Conversely, they drive substantial increases in substantive technical stages, including higher Validation activities (adjusted difference of $+0.285$, 95\% CI $[+0.219, +0.354]$) and Resolution activities (adjusted difference of $+0.212$, 95\% CI $[+0.173, +0.252]$). This indicates that higher completeness allows maintainers to bypass preliminary triage friction and engage directly in deeper technical investigations. 

Furthermore, examinations of report-level validity and specific technical materials reveal a critical nuance: confirming a bug's existence does not equate to endorsing its accompanying technical details. Even within maintainer-confirmed valid reports, a substantial share of detailed artifacts—specifically 32.65\% of technical claims and 33.76\% of proposed fixes—are ultimately classified as rejected or corrected. This demonstrates that while structured reports can successfully flag genuine defects, their granular technical components still fundamentally require rigorous human verification and correction.
\section{Report Length Distributions}
\label{app:report-length}
\begin{figure}[t]
\vspace{-4mm}
\centering
\subfloat[Overall.\label{fig:report-length-overall}]{%
  \includegraphics[width=0.49\columnwidth]{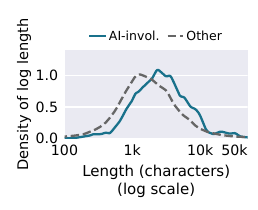}}\hfill
\subfloat[By AI role.\label{fig:report-length-by-role}]{%
  \includegraphics[width=0.49\columnwidth]{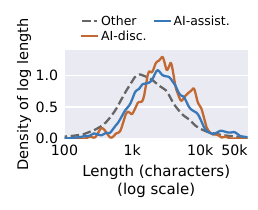}}
\caption{Report length distributions.}
\label{fig:report-length}
\vspace{-2mm}
\end{figure}

\autoref{fig:report-length} compares initial-report lengths, measured in Unicode characters after trimming surrounding whitespace. The count includes Markdown, code, and logs but excludes titles and follow-up comments. AI-involved reports are generally longer, with a median of 2,573 characters versus 1,636 for Other reports. AI-discovered reports have the highest median (2,853.5), followed by AI-assisted reports (2,528.5). Both distributions tend toward longer reports than Other.

\begin{figure}[t]
\centering
\begingroup
\setlength{\fboxsep}{6pt}
\fbox{%
\begin{minipage}{\dimexpr\columnwidth-2\fboxsep-2\fboxrule\relax}
\small
\noindent
\begin{minipage}{\linewidth}
\raggedright
\textbf{(a) Browser-use \#3939}\hfill\textit{Jan.\ 25, 2026}
\par\medskip
\textbf{Reporter:}\par
``Generated by Argus Security v1.0.15 - 6-Phase \textcolor{blue!65!black}{AI-Powered Security Analysis}''
\par\smallskip
``Model: \textcolor{blue!65!black}{Claude Sonnet 4.5} (claude-sonnet-4-5-20250929)''
\par\smallskip
``Total Findings: 10 (after 23.1\% false positive reduction via multi-agent review)''
\par\smallskip
``Attack Scenario (Validated):''\par
\texttt{\# If TEMPLATE\_REPO\_URL is controllable:}
\par\medskip
\textbf{Project responder:}\par
``\textcolor{red!65!black}{These findings aren't accurate}''
\par\smallskip
``SSRF via urlopen: \texttt{TEMPLATE\_REPO\_URL} is a hardcoded constant pointing to GitHub.''
\par\smallskip
\end{minipage}
\par\medskip\hrule\medskip
\begin{minipage}{\linewidth}
\raggedright
\textbf{(b) OpenSSL \#30330}\hfill\textit{March 9, 2026}
\par\medskip
\textbf{Reporter:}\par
``\path{rv64i_zkne_set_encrypt_key} logic for checking for null keys is backwards''
\par\smallskip
``I'm not a RISC-V expert, but I believe the intended logic is what the comment says: if either userKey or key is null (0) return -1.
\par\smallskip
What the code \emph{actually} does is that if either of them are non-null it proceeds (bnez branches on non-zero and then jumps to the actual execution code).''
\par\smallskip
``\textcolor{blue!65!black}{This was found with Claude :-)}''
\par\medskip
\textbf{Maintainer:}\par
``Sorry about that. \textcolor{red!65!black}{I'll fix it in a new PR.} It was a copy-paste of \path{aes-riscv64.pl}:''
\par\smallskip
``\textcolor{red!65!black}{I'll add a test} in the next commit covering all conditions for \path{AES_set_encrypt[/decrypt]_key}.''
\end{minipage}
\end{minipage}}
\endgroup
\caption{Contrasting AI-discovered reports: (a) rejected security claims in Browser-use; (b) an acknowledged null-check defect in OpenSSL. Colors emphasize AI involvement and project responses.}
\label{fig:report-examples}
\end{figure}

\section{Example Cases}
\label{app:examples}

\autoref{fig:report-examples} presents two AI-discovered reports with contrasting maintainer verdicts. Excerpts may span multiple posts, with blue highlighting AI involvement and dark red indicating project responses. Personal names and usernames are omitted, while tool names are retained. Formatting is adapted, with omissions marked.

In case (a), an Argus Security report using Claude Sonnet 4.5 claimed multiple vulnerabilities in Browser-use, including an SSRF scenario involving \texttt{TEMPLATE\_REPO\_URL}. However, the responder clarified that the URL was a hardcoded constant rather than attacker-controlled input, and that Python validates certificates by default. These incorrect assumptions support the \emph{Invalid} verdict, illustrating how detailed AI-generated analysis can rely on flawed premises.

In case (b), Claude identified incorrect null-check logic in the source code. The maintainer confirmed the defect and committed to a fix and additional tests, supporting the \emph{Valid} verdict. Unlike case (a), this report identified a concrete, confirmed logic error.

\section{Definitions of Study Measures}
\label{app:definitions}

Tables~\ref{tab:definitions-rq1}--\ref{tab:definitions-rq3} summarize the definitions used in our LLM annotation prompts. 

\begin{table*}[p]
\centering\footnotesize
\renewcommand{\arraystretch}{0.95}
\caption{Definitions of study measures: Study Scope and AI Use in Bug Reporting.}
\label{tab:definitions-rq1}
\resizebox{0.95\textwidth}{!}{%
\begin{tabular}{@{}p{0.14\textwidth}p{0.82\textwidth}@{}}
\toprule Category & Definition \\
\midrule
\textit{Discovery} & \textbf{Source Code Review}. Discovered by inspecting, reviewing, or auditing source code directly, by a human, LLM/AI agent, or another code-review process; excludes automated static analyzers. \\
 & \textbf{Static Analysis}. Discovered by automated static analysis without executing the target, such as CodeQL, Infer, Coverity, Cppcheck, or static taint analysis. \\
 & \textbf{Fuzzing}. Discovered by fuzzing or automatically generated/mutated inputs, such as AFL, libFuzzer, Syzkaller, or another fuzzing system. \\
 & \textbf{Dynamic Analysis}. Discovered primarily through a runtime bug-detection tool such as AddressSanitizer, UBSan, TSan, MSan, Valgrind, or a similar dynamic analysis tool. \\
 & \textbf{Testing}. Discovered while running unit, regression, self-, integration, or other deliberately designed tests. \\
 & \textbf{Build / Installation}. Discovered during configuration, compilation, linking, packaging, building, or installation. \\
 & \textbf{Operational Usage}. Encountered during normal usage, deployment, production, or another real-world scenario. \\
\midrule
\textit{AI task} & \textbf{Problem Diagnosis}. AI interprets code, logs, or symptoms; explains failure mechanisms; identifies suspected defects, root causes, or regression commits; or assesses existing fixes. \\
 & \textbf{Debugging Support}. AI suggests investigation steps, configures debugging tools, writes or modifies logging/trace/dump scripts, or helps collect diagnostic information. \\
 & \textbf{Reproduction and Testing}. AI creates or modifies reproducers, PoCs, triggering inputs, workloads, diagnostic tests, or regression tests, or helps execute these tests. \\
 & \textbf{Fix Recommendation}. AI proposes a fix or workaround, generates or modifies a patch, or recommends configuration changes to resolve or mitigate the issue. \\
 & \textbf{Information Retrieval}. AI searches for relevant issues, commits, patches, documentation, or similar cases. \\
 & \textbf{Report Writing}. AI drafts, organizes, translates, or edits the report or discussion replies. \\
\bottomrule\end{tabular}%
}
\par\vspace{5pt}
\caption{Definitions of study measures: Report Reliability (RQ2).}
\label{tab:definitions-rq2}
\resizebox{0.95\textwidth}{!}{%
\begin{tabular}{@{}p{0.14\textwidth}p{0.82\textwidth}@{}}
\toprule Category & Definition \\
\midrule
\textit{Report elements} & \textbf{Summary}. A concise description of the abnormal behavior. \\
 & \textbf{Affected component}. The affected module, function, API, file, subsystem, or feature. \\
 & \textbf{Triggering conditions}. The input, condition, configuration, or operation that triggers the problem. \\
 & \textbf{Reproduction steps}. An ordered procedure, command sequence, or code example allowing reproduction. \\
 & \textbf{Observed behavior}. The actual result: an error, crash, incorrect output, timeout, or performance degradation. \\
 & \textbf{Expected behavior}. What should happen under the same conditions. \\
 & \textbf{Environment}. Relevant version and execution context, such as software version, commit, OS, architecture, compiler, dependencies, hardware, or build configuration. \\
\midrule
\textit{Report quality} & \textbf{High}. All elements are concrete enough for investigation without major assumptions. \\
 & \textbf{Medium}. The report is understandable but one or more elements are incomplete, vague, or missing. \\
 & \textbf{Low}. The report is vague, observed behavior is missing or unclear, trigger and reproduction steps are both missing, the component is unidentified, or independent evaluation is infeasible. \\
\midrule
\textit{Rejection reason} & \textbf{Misuse}. The issue results from incorrect configuration or API usage. \\
 & \textbf{Expected Behavior}. The reported behavior is intended or consistent with documented specifications. \\
 & \textbf{Unreproducible}. The issue cannot be reproduced using the provided steps or supporting materials. \\
 & \textbf{Incorrect Technical Claims}. The report makes incorrect claims about APIs, execution paths, root causes, or impact. \\
 & \textbf{Out-of-Scope}. The issue belongs to another project, affects an unsupported version, or occurs only in a development or test environment maintainers consider outside the project's scope. \\
 & \textbf{Others}. An identifiable rejection reason outside the five named categories, such as insufficient information, submission-policy violations, duplicate reports, or an issue already fixed before submission. \\
\bottomrule\end{tabular}%
}
\par\vspace{5pt}
\caption{Definitions of study measures: Maintainer Practices (RQ3).}
\label{tab:definitions-rq3}
\resizebox{0.95\textwidth}{!}{%
\begin{tabular}{@{}p{0.14\textwidth}p{0.82\textwidth}@{}}
\toprule Category & Definition \\
\midrule
\textit{Triage} & \textbf{Information Request}. Asks for logs, environment, version, reproduction steps, examples, stack traces, screenshots, configuration, or other missing information. \\
 & \textbf{Duplicate Checking}. Searches for, mentions, or assesses duplication of an existing report or issue. \\
\midrule
\textit{Validation} & \textbf{Reproduction and Testing}. Tries to reproduce the issue, runs a test or command, reports a reproduction result, or asks the reporter to run an additional test. \\
 & \textbf{Code and API Investigation}. Examines source code, implementation behavior, API semantics, contracts, documentation, stack traces, or technical claims. \\
\midrule
\textit{Resolution} & \textbf{Fix Recommendation}. Proposes or links a fix, patch, diff, PR, commit, code change, plan, or workaround. \\
 & \textbf{Patch Review and Testing}. Reviews a patch/PR/diff, requests or performs regression/unit testing, or verifies that a fix works. \\
\midrule
\textit{Material type} & \textbf{Reproducer}. Code, commands, inputs, or steps for reproducing the bug. \\
 & \textbf{Technical claim}. A specific assertion about the cause, behavior, execution path, or security impact. \\
 & \textbf{Proposed fix}. A patch, concrete fix suggestion, or workaround. \\
\midrule
\textit{Material action} & \textbf{Use}. Explicitly uses the material, such as running the reproducer or testing the proposed patch. \\
 & \textbf{Request supplementation}. Asks for missing information or materials needed to evaluate it. \\
 & \textbf{Reject}. Identifies a specific error and provides or describes a correction, or rejects a specific claim or fix with a technical reason. \\
\bottomrule\end{tabular}%
}
\end{table*}

\end{document}